\documentclass[times,doublespace]{chrisqj}

\usepackage[dvips,colorlinks,bookmarksopen,bookmarksnumbered,citecolor=red,urlcolor=red]{hyperref}

\usepackage{moreverb}

\usepackage[mathscr]{eucal}
\usepackage{amsmath}

\newcommand\um{\mu\mathrm{m}}
\newcommand\nm{\mathrm{nm}}

\def\volumeyear{2009}

\begin{document}

\runningheads{C.~D.~Westbrook \etal}{Specular reflection from ice crystals}

\title{Doppler lidar measurements of oriented planar ice crystals\\ falling from supercooled and glaciated layer clouds}


\author{C.~D.~Westbrook\corrauth, A.~J.~Illingworth, E.~J.~O'Connor and R.~J.~Hogan}

\address{Deptartment of Meteorology, University of Reading, UK}

\corraddr{Dr. Chris Westbrook, Department of Meterology, University of Reading, Earley Gate, Reading, Berkshire, RG6 6BB, UK. E-mail: c.d.westbrook@reading.ac.uk}

\begin{abstract}
The properties of planar ice crystals settling horizontally have been investigated using a vertically-pointing Doppler lidar. Strong specular reflections were observed from their oriented basal facets, identified by comparison with a second lidar pointing $4^{\circ}$ from zenith. Analysis of 17 months of continuous high-resolution observations reveal that these pristine crystals are frequently observed in ice falling from mid-level mixed-phase layer clouds (85\% of the time for layers at $-15^{\circ}$C). Detailed analysis of a case study indicates that the crystals are nucleated and grow rapidly within the supercooled layer, then fall out, forming well-defined layers of specular reflection. From the lidar alone the fraction of oriented crystals cannot be quantified, but polarimetric radar measurements confirmed that a substantial fraction of the crystal population was well oriented. As the crystals fall into subsaturated air, specular reflection is observed to switch off as the crystal faces become rounded and lose their faceted structure.

Specular reflection in ice falling from supercooled layers colder than $-22^{\circ}$C was also observed, but was much less pronounced than at warmer temperatures: we suggest that in cold clouds it is the small droplets in the distribution that freeze into plates and produce specular reflection, whilst larger droplets freeze into complex polycrystals. 

The lidar Doppler measurements show that typical fall speeds for the oriented crystals are $\approx0.3\mathrm{ms^{-1}}$, with a weak temperature correlation; the corresponding Reynolds number is $\mathrm{Re}\sim10$, in agreement with light-pillar measurements. Coincident Doppler radar observations show no correlation between the specular enhancement and eddy dissipation rate, indicating that turbulence does not control crystal orientation in these clouds.

\end{abstract}


\received{5 January 2007}
\revised{\quad}
\accepted{\quad}

\maketitle

\section{Introduction}
It has been recognised for some time that single pristine ice crystals often fall with their major axes horizontal over a certain range of Reynolds number. Polarimetric radar measurements and in-situ sampling led Hogan \etal~(2003a) and Field \etal~(2004) to suggest that concentrations of these pristine crystals were often associated with the presence of mixed-phase cloud layers, in which the crystals were nucleated and grew rapidly by vapour deposition. These supercooled layer clouds are a frequent occurrence in the atmosphere (Hogan \etal~2003b, 2004) but their microphysics is little studied (Fleishauer \etal~2002), and they are poorly simulated in numerical weather models at present (Bodas-Salcedo \etal~2008). 

Vertical lidar measurements are an alternative and powerful means of observing these pristine crystal populations. If the crystals are planar (and given the lack of active ice nuclei at temperatures warmer than $-8^{\circ}$C this is the case for many ice-producing supercooled layers, Pruppacher and Klett 1997) their basal facets will tend to orient in the horizontal plane, and strong `mirror-like' specular reflections are observed as the lidar beam is reflected directly back to the receiver (Platt 1978, Platt \etal~1978). 
Simultaneous measurements at vertical and at a few degrees from zenith allows reliable identification of oriented crystal populations from the large difference in backscatter observed at the two angles (Thomas \etal~1990). In this paper we apply this technique to identify layers of planar crystals, and investigate some of their characteristics using a vertically pointing Doppler lidar at the Chilbolton Observatory in Hampshire. This instrument has operated continuously since September 2006, and the extensive data set allows us to provide a much more thorough statistical investigation of the conditions in which these crystals occur than has previously been possible, as well as providing many fascinating case studies. These are (to the authors' knowledge) the first measurements of Doppler velocities from specularly reflecting crystals, and this new information provides valuable insight into the fall speeds and likely sizes of the crystals involved.

The article comprises a review of previous observations and theory on crystal orientation and specular reflection, followed by details of the instruments used, a case study of a supercooled layer cloud precipitating oriented pristine crystals, and a statistical analysis of 17 months of continuous lidar data. A brief discussion concludes the paper.

\section{Review of previous observations and theory}
Although the orientation of ice crystals as they fall is not yet fully understood, we can identify some general features. Very small crystals $\sim10\um$ in diameter are affected by rotational Brownian motion (thermal noise) and have an approximately random orientation as they fall. Drag is produced as vorticity generated at the crystal surface diffuses away (Batchelor 1967), and orienting torques are weak (Katz 1998): these crystals are unlikely to produce significant specular reflection. As the crystals grow bigger and the Reynolds number $\mathrm{Re}=vD/\nu_k$ increases to $\sim1$ ($D$ here is the diameter of the crystal, $v$ its fall speed, and $\nu_k$ the kinematic viscosity of the air), the time taken to fall a distance equal to the particle diameter $t=D/v$ becomes comparable to the time taken for the vorticity generated at the crystal surface to diffuse that same distance $t\approx D^2/\nu_k$, and the vorticity is `left behind' as it falls. Stable, symmetrical vortices are formed behind falling discs (Willmarth \etal~1964, Pitter \etal~1973); numerical experiments by Wang and Ji (1997) confirm that these standing eddies also form behind hexagonal plate and broad-branch ice crystal shapes. If the crystal is inclined at an angle to the horizontal plane the asymmetrical drag acts to reorient the plate into the horizontal position: this reorientation is stabilised by viscosity, so oscillations around the horizontal are damped, and the particle falls stably and steadily (Willmarth \etal~1964). Measurements of natural ice crystals by Kajikawa (1992) confirm that horizontal orientation occurs for a variety of planar crystal types (all of the P1 and P2 types defined by Magono and Lee 1966). 
As the crystals grow even larger, the inertial forces are too strong to be damped, and the horizontal orientation becomes unstable: this initially affects the wake downstream producing a slow pitching motion, whilst at larger Reynolds number the vortices begin to shed fluid behind the falling crystal leading to more irregular tumbling motions (Willmarth \etal~1964, Kajikawa 1992). Circular discs fall steadily between $1<\mathrm{Re}<100$; for the natural ice crystals measured by Kajikawa the maximum Reynolds number for stable orientation was in the range $\mathrm{Re}\simeq50$--$100$, and depended on the crystal habit. 

Provided that enough crystals are falling in the preferred Reynolds number regime, and they have sufficient faceted surface, strong specular reflection is possible. We note that even branched and dendritic planar crystals typically have a faceted plate at their centre (eg. Bentley and Humphreys 1964) which may allow specular reflection to occur. Platt (1977) first observed layers of unusually strong backscatter in a mixed-phase altostratus cloud. Model calculations (Platt 1978) for populations of ice discs with various flutter angles showed that specular reflection was able to explain the vertical lidar observations. To clinch the interpretation, Platt \etal~(1978) made scanning observations in a thin mid-level ice cloud where very strong backscatter was observed at vertical with almost almost no depolarisation (specular reflection at normal incidence does not rotate the plane of polarised light); scanning the lidar $2^{\circ}$ away from vertical decreased the measured backscatter by an order of magnitude, and increased the depolarisation ratio to 25\%. At the same time, a sub-sun (similar to a light pillar, see below) was reported by a research aircraft flying above, confirming the presence of specularly reflecting crystals. 

Since then, a number of similar observations have been made. Sassen (1984) observed strong backscatter and low depolarisation in the ice falling beneath a supercooled liquid layer embedded in a deep orographic cloud. Platt \etal~(1987) identified layers of ice cloud with anomalously large backscatter and low depolarisation (integrated through the layer). Similarly, Hogan and Illingworth (2003) used integrated backscatter measurements to identify regions of strong specular reflection, and found it to occur exclusively between $-10$ and $-20^{\circ}$C for 20 days of lidar data. Thomas \etal~(1990) stepped between $0$ and $3^{\circ}$ from zenith to identify regions of cirrus affected by specular reflection. Collating statistics from cirrus cases over 3 years they found specular reflection was present approximately half the time at all heights. They also scanned in 1mrad intervals around zenith in one case study, and inferred from the fall-off in backscatter that the angular wobble of the crystals was approximately $0.3^{\circ}$. Young \etal~(2000) observed 8 mid-level mixed-phase clouds and found specular reflection in 20\% of layers between $-10$ and $-20^{\circ}$C. Sassen and Benson (2001) analysed long term statistics of depolarisation ratio in ice cloud, and found a minimum in the range $-10$ to $-20^{\circ}$C which was not present in their analysis of off-zenith measurements, indicating the frequent occurrence of specular reflection at these temperatures. Noel and Sassen (2005) used scanning lidar measurements of backscatter and depolarisation and compared them to forward modelled values to estimate the likely range of crystal flutter; their retrieved angular distribution half-widths were typically $1$--$2^{\circ}$. From space Hu (2007) analysed returns from the CALIPSO lidar which initially pointed $0.3^{\circ}$ from nadir, and estimated that half of all optically thick ice cloud was affected by specular reflection, with important repercussions for retrievals of cloud optical properties. The statistics from these studies are summarised in table \ref{otherstudies}.
\begin{table*}[t]
\caption{\label{otherstudies}Previous statistics of specular reflection}
\centering
\begin{tabsize}
\begin{tabular}{lll}
\toprule
Study&Sample&Results\\
\midrule
Platt \etal~(1987)&48 mid-latitude cirrus cases & 42\% of profiles with $-20>T_{m}>-30^{\circ}$C affected,\\
 & &20\% of profiles with $-30>T_{m}>-40^{\circ}$C\\
 & +11 tropical cases &Trop. cases: no specular reflection observed.\\
 &&(nb. $T_{m}=$ mid-cloud temperature)\\

Thomas \etal~(1990)&Accumulated mid-latitude &Present in approximately 50\% of cloud at all heights\\
 & cirrus cases from 3 years&most of the observations clustered around 9~km altitude\\

Young \etal~(2000)&8 mid-latitude case studies&Observed 20\% of time in layers $-10>T_{m}>-20^{\circ}$C,\\
 &mid-level, mixed-phase &did not occur at other temperatures\\

Hogan and Illingworth (2003)&20 days continuous data&Observed 18\% of time in cloud layers at 4~km,\\
& &not observed at cirrus altitudes\\

Noel and Sassen (2005)&6 cirrus cases&Observed 22\% of time for $T>-20^{\circ}$C,\\
& &35\% in cold clouds $T<-30^{\circ}$C\\

Hu (2007)&1 month of CALIPSO data&Half of all optically thick ice cloud profiles affected\\

Chepfer \etal~(1999)&3 months POLDER data&37-50\% of all ice-only cloud show specular glint\\

Noel and Chepfer \etal~(2004)&POLDER, 31 ice cloud cases&80\% of ice-only cloud showed specular glint\\

\bottomrule
\end{tabular}
\end{tabsize}
\end{table*}

A number of theoretical calculations have been performed to estimate the specular backscatter from a population of oriented crystals (Platt 1978, Popov and Shefer 1994, Iwasaki and Okamoto 2001); however knowledge of the amount of faceted surface on the crystal, and distribution of crystal orientations is required, and these quantities are not well constrained at present. As an example, to produce a backscatter of $10^{-4}\mathrm{m^{-1}sr^{-1}}$ only requires a concentration of $\sim1\mathrm{m^{-3}}$ ice discs $500\um$ in diameter if they are all perfectly aligned. However in practice a distribution of crystal orientations must always exist, and the above observations show that it will be much wider than the lidar field of view ($66\mu\mathrm{rad}$ for our monostatic Doppler lidar). This means that the true planar crystal population is much larger ($1000\times$ larger for a uniform $2^{\circ}$ flutter), and only a small fraction of the crystals are reflecting the lidar beam back to the receiver at any one time. Likewise, if only part of the crystal surface is faceted, a correspondingly larger crystal population is required. Linking crystal concentrations to observed backscatter values is therefore very difficult.



In addition to lidar measurements, passive measurements of radiances from space can also be affected as sunlight glints off oriented crystals: Chepfer \etal~(1999) analysed 30 days of POLDER data and found that this happened in at least 40\% of ice cloud. Note this instrument has a much wider field of view than lidar (6.7km pixel size). Noel and Chepfer (2004) looked at 31 ice cloud cases, and found 80\% were affected by a specular reflection. Looking at the change in radiance around the glint direction and comparing to simulated reflectances, they estimated around 10\% of crystals were horizontally oriented on average. Most recently, Br\'{e}on and Dubrulle (2004) used a crystal orientation model, combined with radiative transfer calculations and estimated that the fraction of oriented crystals was typically only 1\%. Clouds containing supercooled liquid were excluded from these studies by detection of the rainbow scattering pattern from spherical liquid droplets.

Finally, Sassen (1980) took photographs of light pillars produced by artificial lights during snowfall. These optical phenomena are produced by oriented crystals between the light source and observer, which reflect the light from their horizontal facets, producing a pillar of light apparently extending vertically from the light source. Any crystal wobble produces a spreading out of the pillar, as rays of light out of the plane formed by the light pillar, the light source and the observer are also reflected (to the observer), leading to a finite pillar width. By measuring this pillar width, Sassen estimated the characteristic wobble of the crystals in the snowfall, with observed values in the range $0.5$--$3^{\circ}$. Correlating this width with the mean size of crystals (mostly plate and branched planar crystals) collected at the ground, and using empirical velocity-diameter relationships, his data showed a minimum in the pillar width (and therefore crystal wobble) at a Reynolds number of $\mathrm{Re}\approx8$.





\section{Instruments and method}

The key instrument in this study is a newly developed $1.5\um$ Doppler lidar (HALO Photonics Ltd., Malvern, UK). The instrument measures profiles of backscatter and mean Doppler velocity resolved over 36~metre range gates and recorded every 30~seconds to a range of 10~km.  The lidar has operated continuously since September 2006, and points directly at vertical. Specifications are given in table \ref{specifications}. In addition to this instrument, a $905\nm$ lidar ceilometer records backscatter profiles at approximately the same range and temporal resolution (Illingworth \etal~2007), and points $4^{\circ}$ off vertical. This lidar is calibrated using measurements of the integrated backscatter in optically thick stratocumulus cloud as described by O'Connor \etal~(2004). Mie calculations for liquid water droplets indicate that the $1.5\um$ lidar can be calibrated in a similar way: details of this are given in Appendix A.

\begin{table}
\caption{\label{specifications}Specifications of Doppler Lidar}
\centering
\begin{tabsize}
\begin{tabular}{ll}
\toprule
Wavelength&$1.5~\um$\\
Range resolution&36~m\\
Integration time&30~s\\
Maximum range&10~km\\
Maximum Doppler velocity&10~m/s\\
Doppler velocity noise&$\sim0.05~$m/s (SNR dependent)\\
Pulse duration&200~ns\\
PRF&15~kHz\\
Beamwidth&60~mm (at source)\\
Divergence&$33~\mu$rad\\
Antenna&monostatic optic-fibre coupled\\
\bottomrule
\end{tabular}
\end{tabsize}
\end{table}

Radar measurements were made using 35- and 94-GHz Doppler radars: both instruments are described by Illingworth \etal~(2007). The 35GHz radar provides reflectivity (at two orthogonal polarisations) and Doppler parameters with a 30~second integration time, and has a gate length of 30~metres. This instrument was operated at $45^{\circ}$ elevation for the case study in section 4, and at vertical for the statistics in section 5. The 94~GHz radar operates on a case study basis to preserve the lifetime of the tube, measuring reflectivity, and Doppler parameters at 1~second resolution, with a gate length of 60~metres.  Full Doppler spectra are also recorded once per minute by both radars when cloud is present; the integration time for these spectra is 1 second. 

A key parameter in this study is the lidar `colour ratio' which we define as the ratio of the lidar backscatter measured by the $905\mathrm{nm}$ ceilometer to that measured by the $1.5\um$ Doppler lidar:
\begin{equation}
\textrm{Colour ratio [dB]}=10\log_{10}\left[\frac{\beta_{905\mathrm{nm}}}{\beta_{1.5\um}}\right]
\end{equation}
Under conditions where there is no specular reflection we expect this ratio to be greater than $0\mathrm{dB}$ for ice particles, since the imaginary component of the refractive index of ice is 3 orders of magnitude larger at $1.5\um$ compared to $905\nm$ ($n''=4.7\times10^{-4}$ vs. $n''=4.3\times10^{-7}$, Warren 1984). This results in significant absorption of photons as they travel through the ice crystal, and a corresponding reduction in the backscatter\footnote{We note that the attenuation of the lidar beam as it passes through a unit volume of cloud depends only the extinction of the ice particles within the cloud, which should be the same at both wavelengths (equal to twice the projected area of the crystals per units volume of air, assuming geometric optics). As a result the ratio of the measured attenuated backscatters should be identical to the ratio of the true (unattenuated) backscatters, provided that the ice cloud sampled by the two lidar beams is the same (ie. it is horizontally homogeneous on scales of a few hundred metres). We have neglected attenuation by aerosol particles, and multiple scattering between ice crystals.}. As a guide to the magnitude of this effect, consider a ray of light as it undergoes a series of internal reflections through the ice crystal and back to the lidar telescope. According to the Beer-Lambert rule, its intensity will be reduced by a factor $\exp(-4\pi n'' x/\lambda)$, so for a total path length of $x=200\um$, the backscatter at $\lambda=1.5\um$ is reduced by $3.4$dB, whilst light at the ceilometer wavelength is essentially unaffected. We conclude that this differential absorption is likely to be significant for ice crystals in the atmosphere which are typically a few hundred microns to a few millimetres in size, and expect positive colour ratios $>0\mathrm{dB}$. This effect could in principle be used to measure crystal size in ice clouds, and will be explored in a separate publication.

\begin{figure}
\centering
\includegraphics[width=3in]{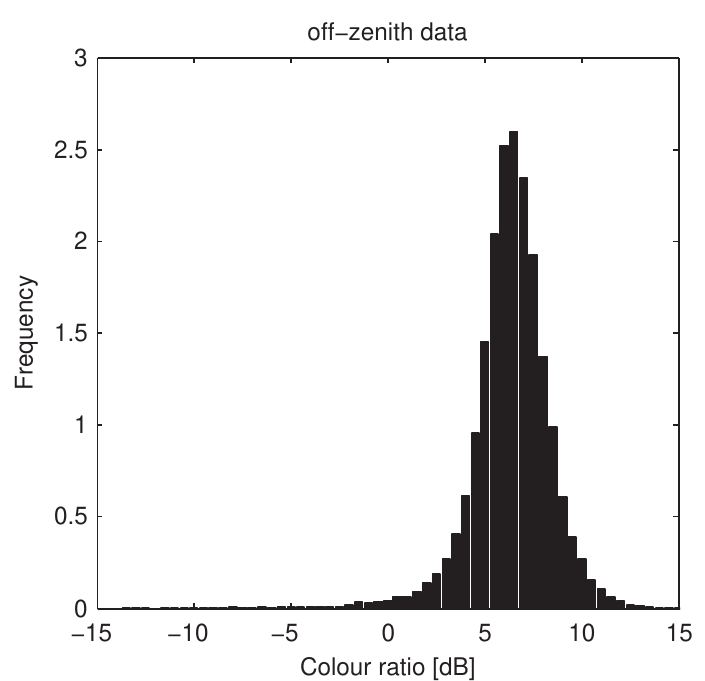}
\caption{\label{OZ}Distribution of colour ratios ($905\nm/1.5\um$ backscatter) collected in ice cloud while both lidars were pointing off-zenith. Frequencies are normalised such that the total area under the curve =10.}
\end{figure}
To check the analysis above is correct, we have collected 9 days of backscatter data with both lidars pointing off-zenith. During this period (25 February to 4 March 2008) various types of ice cloud were observed including cirrus, altocumulus with ice fall streaks below, and thick frontal ice cloud. Quicklooks of the lidar data may be viewed at \underline{www.met.reading.ac.uk/radar/realtime/}. From this data figure \ref{OZ} shows the histogram of the measured colour ratio in the ice-phase regions. It is apparent that the absorption effect is significant, with 95\% of the data corresponding to colour ratios between $+3$ and $+12$~dB (ie. backscatter at $1.5\um$ reduced by a factor  2 to 16). Almost no data were observed with colour ratios $<0\mathrm{dB}$ --- the 1\% of pixels that were, revealed themselves on closer inspection to be isolated points resulting from occasional artefacts recorded by the ceilometer.

These off-vertical measurements contrast sharply with our observations with the $1.5\um$ Doppler lidar pointing directly vertical. In such cases we frequently observe colour ratios $<0$~dB, in some cases as low as $-25$~dB (ie. $300\times$ more backscatter at vertical). We conclude that these negative colour ratios must correspond to cases where the ice crystals are highly aligned in the horizontal, ie. the specular reflection from the horizontal crystal facets produces a strong signal in the vertical Doppler lidar data which is not present in the off-vertical ceilometer data, leading to the negative colour ratios. In what follows we will use this condition to identify regions of cloud where specular reflection is occurring (negative colour ratios) and where it is not (positive colour ratios). 

%

\section{Case study, 18 May 2008}
\subsection{Vertical measurements}
\begin{figure*}
\centering
\includegraphics[width=7in]{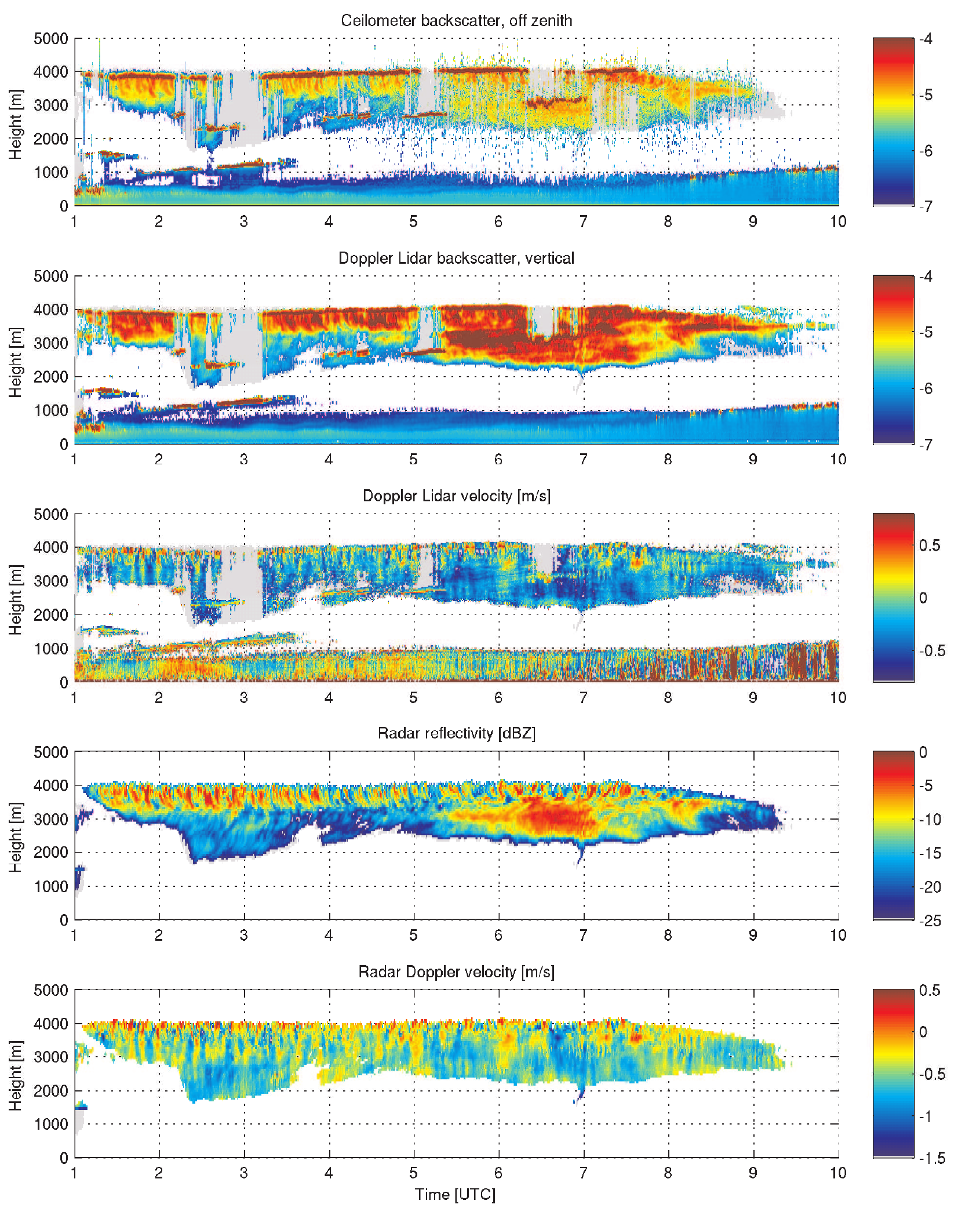}
\caption{\label{7lidars}Lidar and radar observations, morning of 18 May 2008. Top panel shows the backscatter measured by the ceilometer pointing at $4^{\circ}$ from zenith; second panel shows the backscatter measured by the Doppler lidar pointing directly at zenith. Colour scale is in logarithmic units ie. $\log_{10}(\mathrm{m^{-1}sr^{-1}})$. Third panel shows the vertical velocity measured by the Doppler lidar (negative velocities correspond to downward motion). Final two panels shows the 94-GHz radar reflectivity and vertical Doppler velocity.}
\end{figure*}

Here we present an example of a persistent mid-level mixed-phase cloud associated with a weak occlusion moving southwest across England on the 17 and 18 May 2008. Patchy mid-level cloud cover was first observed over Chilbolton by the vertically pointing 94GHz cloud radar at approximately 06 UTC on the $17^{\mathrm{th}}$; by 15 UTC this had become a well defined layer, with cloud top at 4000m altitude. At the same time low-level stratocumulus was obscuring the mid-level cloud from view of the lidars; in addition some of the ice fall streaks appeared to be seeding the stratocumulus, producing drizzle. However, by early morning on the 18 May the low cloud began to break up allowing an unobscured view of the mixed-phase cloud above: figure \ref{7lidars} shows the time series from the ceilometer and Doppler lidar for the period 01-10 UTC. The off-zenith ceilometer clearly picks out the location of the thin supercooled liquid water layers, with a well defined primary liquid layer at the top (3800-4100m altitude), and intermittent secondary liquid layers embedded in the 1-2km deep ice fall streaks. The concentration of ice crystals is much lower than the concentration of cloud droplets, and so the backscatter is an order of magnitude smaller in the ice virga than in the supercooled liquid layer; the fall streak structure of the sedimenting ice crystals is clearly visible here. In the lowest 1km weak returns from the boundary-layer aerosol are visible in the lidar time series; the radar meanwhile is insensitive to these tiny aerosol particles.

\begin{figure}
\centering
\includegraphics[width=3.25in]{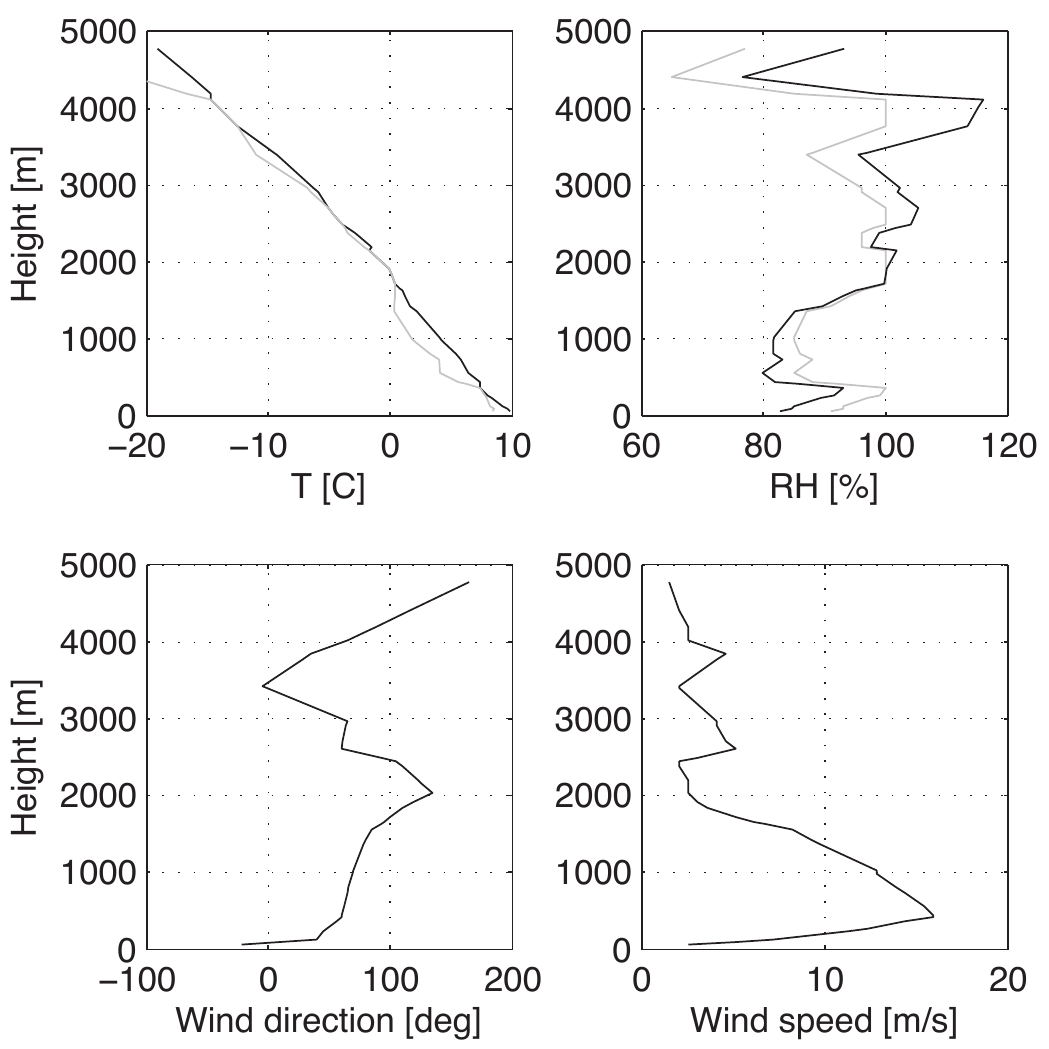}
\caption{\label{larkhillAc}Herstmonceux ascent at 0~UTC on 18 May 2008. Panels clockwise from top left show: air temperature (black) and dew-point (grey); relative humidity with respect to liquid (grey) and ice (black); wind speed; wind direction in degrees.}
\end{figure}

The 0Z sounding from Herstmonceux (125km east of Chilbolton) is shown in figure \ref{larkhillAc}, and the water-saturated layers at 3800-4100m and 2500-2700m seem to be consistent with the ceilometer imagery. The temperature at the top of the uppermost liquid layer was $-15^{\circ}$C; a slight temperature inversion was present at cloud top, and the air above was much drier. Wind speeds were relatively light; directional shear was observed in the ice virga, and is also apparent from the shape of the fall streaks in the radar time series, figure \ref{7lidars}).

Data from the vertically-pointing Doppler lidar are also shown in figure \ref{7lidars}, and it is immediately apparent that the backscatter in the ice virga is typically one order of magnitude (10~dB) larger than that recorded by the ceilometer, in spite of the increased absorption at $1.5\mu$m. This convinces us that oriented planar crystals are present in significant numbers in the virga. These crystals must be nucleated in the primary supercooled layer, growing by deposition in the vapour-rich conditions. The temperature in this water-saturated layer was between $-12.5$ and $-15^{\circ}$C leading to a large supersaturation over ice (see figure \ref{larkhillAc}); Ryan \textit{et al} (1976) found skeletal plates (at $-13$) and stellar crystals (at $-15$) grew rapidly in such conditions. The aspect ratio of their crystals was measured to be more than 10:1 by the time they had reached $100\mu$m in size, and given the discussion above this should lead to a strong reorienting torque (indeed Ryan \textit{et al} noted that nearly all of the crystals settled flat-on as they fell onto the sampling slide). Because of the large vapour excess, the crystals do not grow in such a controlled way as simple hexagonal plates; however there is still likely to be ample faceted surface for specular reflection to occur from.

Unfortunately the microwave radiometers at Chilbolton were not operating on this day. However Korolev \etal~(2007) have developed a correlation between the depth of mid-latitude supercooled layers and their liquid water path from aircraft data, and from this we estimate a value of only $\approx40\mathrm{gm^{-2}}$; this low value is consistent with previous measurements of similar supercooled clouds  (eg. Hogan \etal~2003a,b). As a result riming is likely to be very light or not at all, and the observed strong specular reflection supports this assertion.

In contrast to the lidars, the vertically pointing 94GHz radar (figure \ref{7lidars}) is insensitive to the supercooled liquid droplets in the mixed-phase layer, and is dominated by returns from the much larger ice crystals; again fall streaks are clearly visible. The reflectivity of the ice crystals falling from the primary supercooled layer is observed to vary in the range -15 to 0~dBZ, suggesting ice water contents of $\approx0.01$--$0.1\mathrm{gm^{-3}}$ (Hogan \textit{et al} 2006); this range is consistent with in-situ observations of mid-latitude mixed-phase layer clouds (Carey \textit{et al} 2008).

\begin{figure*}
\centering
\includegraphics[width=7in]{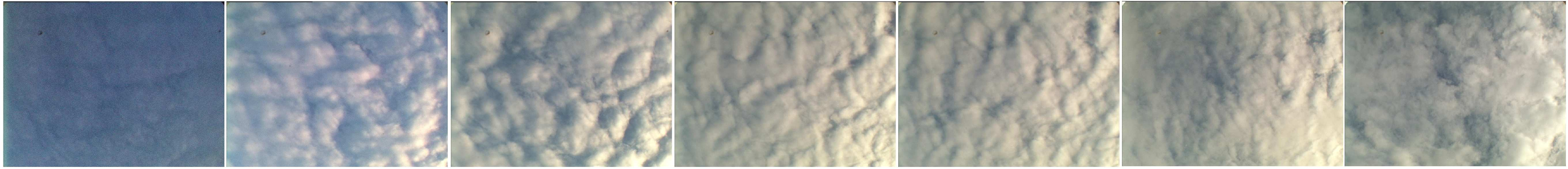}
\caption{\label{sky}Photographs of the sky every 30 minutes from 0420 to 0650 UTC.}
\end{figure*}

Figure \ref{7lidars} shows the vertical velocity from the liquid droplets and ice particles as measured by the Doppler lidar. We have calculated the average Doppler velocity in the uppermost liquid layer for the period shown in figure \ref{7lidars}: this is approximately $+0.02\mathrm{ms^{-1}}$, indicating that any large scale ascent is very gentle. However it is clear that there are smaller scale vertical air motions present - at cloud top there is convective overturning with peak-to-peak vertical motions of $\approx0.5\mathrm{ms^{-1}}$ as the cloud cools radiatively to space. Photographs of the sky in figure \ref{sky} confirm the cellular structure. Waves (approximately 10 minute period) are also present - these are most visible in the earliest part of the time series, and also in the vertical radar Doppler velocity, also shown in figure \ref{7lidars}.

To get an estimate of the terminal velocities of the oriented ice crystals, we have isolated areas with colour ratio $<0\mathrm{dB}$ and calculated the mean lidar  Doppler velocity over the 9 hour period. By averaging over several hours we expect to remove the mesoscale air motions observed in the detailed time series and obtain approximate values for the still-air velocities of the crystals. We estimate this to be $-0.23\mathrm{ms^{-1}}$. Sassen's (1980) light pillar data indicated that there is likely to be a Reynolds number ($\mathrm{Re}\approx8$) where the crystal flutter is minimised, and we expect that these crystals will provide the strongest weighting to the Doppler velocity measurements. Heymsfield and Kajikawa (1987) measured the fall velocity of planar ice crystals collected at the ground in northern Japan (1~km above sea level), and formulated velocity-diameter relationships from those observations. Using the pressure correction method of Beard (1980) we have used this data to calculate  the ranges of Reynolds number and crystal diameter corresponding to a fall speed of $0.23\mathrm{ms^{-1}}$ for various assumed crystal habits: these values are tabulated in (\ref{vre}). We find $\mathrm{Re}\approx4$, $D\approx400\mu$m for the most compact habit (hexagonal plates) increasing to $\mathrm{Re}\approx11$, $D\approx1200\mu$m for the most open (stellar crystals). This range of Reynolds numbers is broadly consistent with Sassen's observations, but may be more robust since we are measuring velocities directly rather than median crystal diameters at the ground (this may be significant since the distribution of crystal orientations is not the same for all crystal diameters).  Our observations are also consistent with Kajikawa's (1992) observations since the Reynolds numbers estimated here are smaller than the critical values where the fall motion becomes unstable.
\begin{table}
\caption{\label{vre}Estimated Reynolds number and crystal diameter corresponding to fall velocity of $0.23\mathrm{ms^{-1}}$}
\centering
\begin{tabsize}
\begin{tabular}{lll}
\toprule
Assumed habit&Re&D [$\mu$m]\\
\midrule
Hexagonal Plates&4&400\\
Sector Plates&5&550\\
Broad-branch&5&550\\
Plates w/extensions&6&650\\
Dendritic&9&1000\\
Stellar&11&1200\\
\bottomrule
\end{tabular}
\end{tabsize}
\end{table}

\begin{figure}
\centering
\includegraphics[width=3.25in]{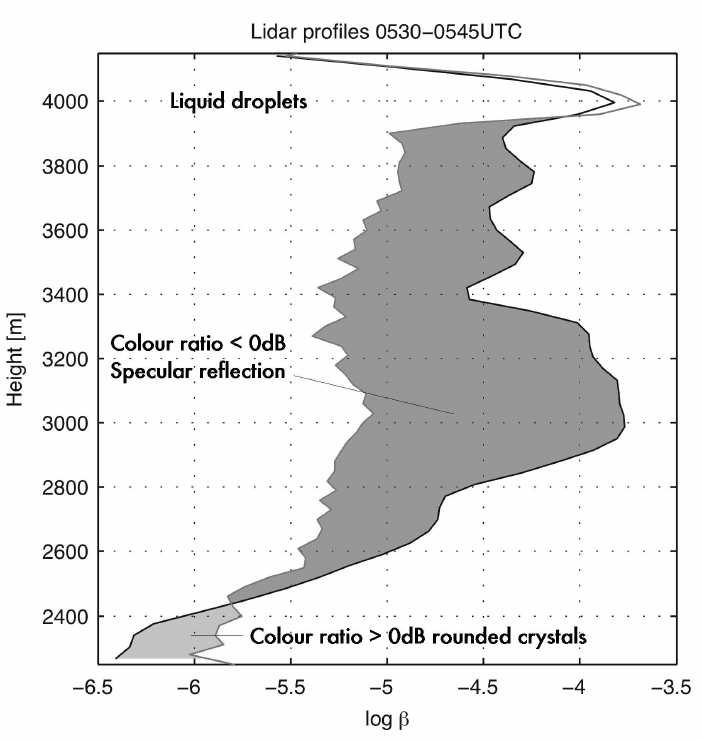}
\caption{\label{7colour}Averaged lidar backscatter profiles for the period 0530-0545~UTC. Grey line is the profile from the off-zenith ceilometer, black line is the profile from the vertically pointing Doppler lidar.  Shading indicates the difference in backscatter measured by the two instruments: dark grey shading corresponds to specular reflection from oriented crystals; light grey shading corresponds to evaporating crystals which have lost their faceted surfaces.}
\end{figure}

Figure \ref{7colour} shows the linearly averaged lidar backscatter profile from the ceilometer and Doppler lidar between 0530 and 0545~UTC (a period when the cloud appears to be relatively steady-state and horizontally homogeneous). The profiles reveal that that specular reflection is maximised at approximately 3000--3300m (colour ratio $=-15$dB), decreasing steadily at lower altitudes. The backscatter profiles in the supercooled liquid are well matched, with slightly higher returns from the ceilometer due to the difference in backscatter to extinction ratio between the two wavelengths (see appendix A). This is evidence that the lidars are well calibrated, and that the attenuation through the ice virga is the same for both lidar beams, despite the enormous difference in backscatter.

The regions near cloud base with positive colour ratios ($\approx+4$~dB) are noted with interest. The fall off in lidar and radar backscatter in this region indicates that it is subsaturated with respect to ice. In these conditions evaporation tends to proceed from crystal edges since this is where the molecules are most weakly bound. If the subsaturation is very slight, evaporation occurs slowly in discrete layers, allowing the crystal to maintain its faceted structure (Hallett \etal~2001). If the crystal falls into drier air however, multiple steps form at the edges (Mason 1971) and the evaporation rate becomes limited by the diffusion of the molecules through the air, leading to strong vapour gradients at the edges and corners (Nelson 1998, Westbrook \etal~2008), producing smooth, rounded crystals (Nelson 1998). Rounding also appears to be more favourable at warmer temperatures (Hallett \etal~2001). Given these results it seems clear that the regions of positive colour ratios are cloud where evaporating crystals have lost their faceted surfaces and become rounded, so specular reflection can no longer occur\footnote{This idea also appears to explain the vertical lidar observations of two mid-level clouds by Uchino \etal~(1988) who were puzzled when they found that that whilst the returns from the middle of their clouds were characteristic of specular reflection (very high backscatter, very low depolarisation ratios), the returns from the base of the virga were characteristic of normal ice cloud (much lower backscatter, depolarisation ratios increased to $\approx30\%$), and that this was co-located with a rapid decrease in relative humidity with respect to ice.} . This is consistent with calculations by Mishchenko \etal~(1997) who showed that flat-faced discs could produce strong specular reflection, whilst thin oblate spheroids (ie. slightly rounded faces) could not. 

An alternative explanation might be that the crystals have evaporated to the point where the crystals are too small to be horizontally oriented (tens of microns in diameter, see section 2). However, given that the radar and lidar Doppler velocities in this region are consistent with crystals several hundred microns in size, we discount this idea.

Identification of these regions where specular reflection is `switched off' may therefore be a useful observational tool to determine if the crystals are evaporating in layers or losing their shape. The latter case is also evidence that the whole crystal surface is close to ice saturation (Nelson 1998), and that the crystal evaporation rates are governed by the standard electrostatic approximation (Mason 1971, Westbrook \etal~2008). 

\begin{figure}
\centering
\includegraphics[width=3.3in]{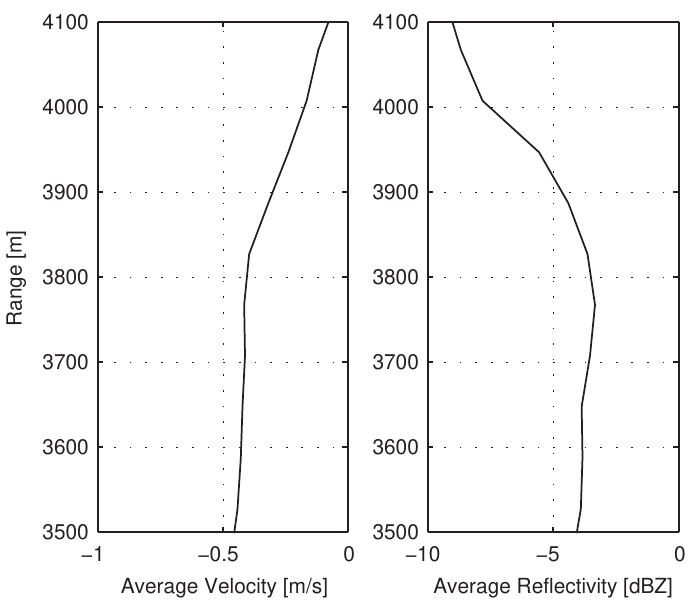}
\caption{\label{testzv}Averaged radar reflectivity and velocity profiles for the period 01-10~UTC for the top 900m of cloud.}
\end{figure}

To investigate the initial growth of the ice at the top the cloud we have averaged the vertical 94-GHz radar Doppler velocity and reflectivity over the whole period (01-10~UTC) for the top 600m, as shown in figure \ref{testzv}. At this range of altitudes the radar data suggest that there is a relatively consistent production of ice. The averaged profiles show that the reflectivity and Doppler velocity of the ice particles increase steadily in magitude through the supercooled layer (the top 300m) reaching values of $\approx-3$dBZ and $-0.45\mathrm{ms^{-1}}$ respectively. Note that the radar velocities are larger than measured by the Doppler lidar in the specular ice, reflecting the weighting of the radar towards the heaviest ice particles. The crystal diameter corresponding to this velocity is rather sensitive to the assumed habit, ranging from $800\mu$m for hexagonal plates to $3500\mu$m assuming stellars (Heymsfield and Kajikawa 1987). We note that over the depth of the supercooled layer the reflectivity increases by about 6dB, which corresponds to a doubling of the average crystal mass over that distance (since reflectivity is, to first order, proportional to mass$^2$) whilst Ryan \etal~(1976) observed crystal masses to increase by a factor of ten between 50 and 150 seconds of growth. The comparison is not so clear-cut because of non-Rayleigh scattering effects and sedimentation, but the implication is that the crystals are not simply nucleated at cloud top and grow whilst steadily falling falling through the water saturated layer, but are also mixed through the layer as the air is gently turned over, smoothing out some of the reflectivity structure. It may be that crystals are also nucleated throughout the depth of the liquid layer rather than only at the top, which would have a similar effect.  Below 3800m there appears to be relatively little change in reflectivity or velocity, suggesting little growth is occuring -- presumably the relative humidity falls off sharply to near ice saturation here (there is some evidence for this in the radiosonde profile, figure \ref{larkhillAc}). 

Unlike Shupe \etal~(2004) we were unable to detect a signature from the supercooled liquid droplets in the radar Doppler spectra (not shown): a single mode was observed, corresponding to the sedimenting ice crystals. Most likely this is because the liquid water content in this cloud is too low to have a detectable radar signature: the case studied by Shupe \etal~had a peak liquid water path of $500\mathrm{gm^{-2}}$ in a 1km thick layer with droplets ascending at $1\mathrm{ms^{-1}}$; our example is much thinner and more weakly forced, and our estimated liquid water path is an order of magnitude smaller. 

\begin{figure}
\centering
\includegraphics[width=3.5in]{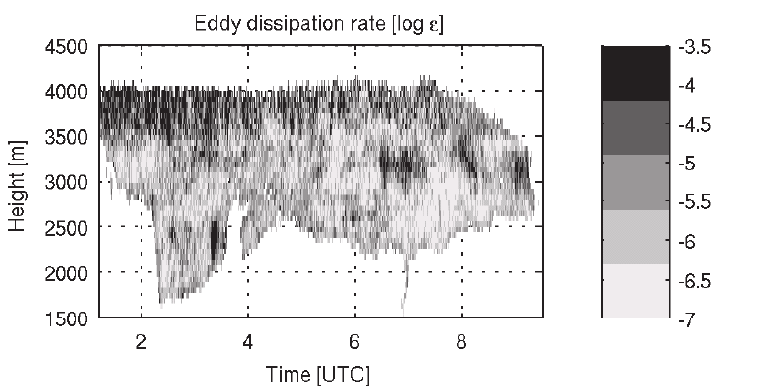}
\caption{\label{teste}Eddy dissipation rate $\epsilon$ estimated from the standard deviation of 1-s mean radar Doppler velocities over 30s periods. Grey scale is in logarithmic units, ie. $\log_{10}(\mathrm{m^2s^{-3}})$.}
\end{figure}

Figure \ref{teste} shows the turbulent eddy dissipation rate $\epsilon$, estimated from the vertically pointing radar as described by Bouniol \etal~(2003). The standard deviation of 1-s snapshots of mean Doppler velocity over 30s periods are combined with wind speed estimates (from the Met Office 12km model forecast), and $\epsilon$ is inferred from the variability in the vertical velocity over the 30s sample time. In most of the virga the air appears to be quite still with dissipation rates of order $10^{-7}$ -- $10^{-6}\mathrm{m^{2}s^{-3}}$, including the evaporation zone at the base of the virga; however in the top 500m of the cloud, this peaks at $\epsilon\approx3\times10^{-4}\mathrm{m^{2}s^{-3}}$, presumably reflecting the radiative cooling of the liquid droplets and associated overturning. This turbulent mixing seems to extend down into the top few hundred metres of the virga; despite this the lidar data indicates that strong specular reflection is still occurring in this region. This indicates that turbulence is unlikely to be a strong influence on crystal orientation, and this is consistent with previous theoretical analysis (Klett 1995): for $\epsilon=3\times10^{-4}\mathrm{m^{2}s^{-3}}$ the Kolmogorov microscale (the size of the smallest eddies) is 1.8mm, and the corresponding velocity perturbation across a 1mm ice crystal is of order $0.005\mathrm{ms^{-1}}$, much smaller than its fall speed. Interestingly, as well as the turbulence in the liquid layer at the top, there are also regions of enhanced $\epsilon$ co-located with the more intermittent supercooled layers embedded in the ice virga (eg. 0630--0700~UTC, 3200m), perhaps associated with the flow of latent heat as these layers are seeded by the crystals falling from above.

\subsection{Polarimetric radar measurements at $45^{\circ}$}

\begin{figure*}
\centering
\includegraphics[width=7in]{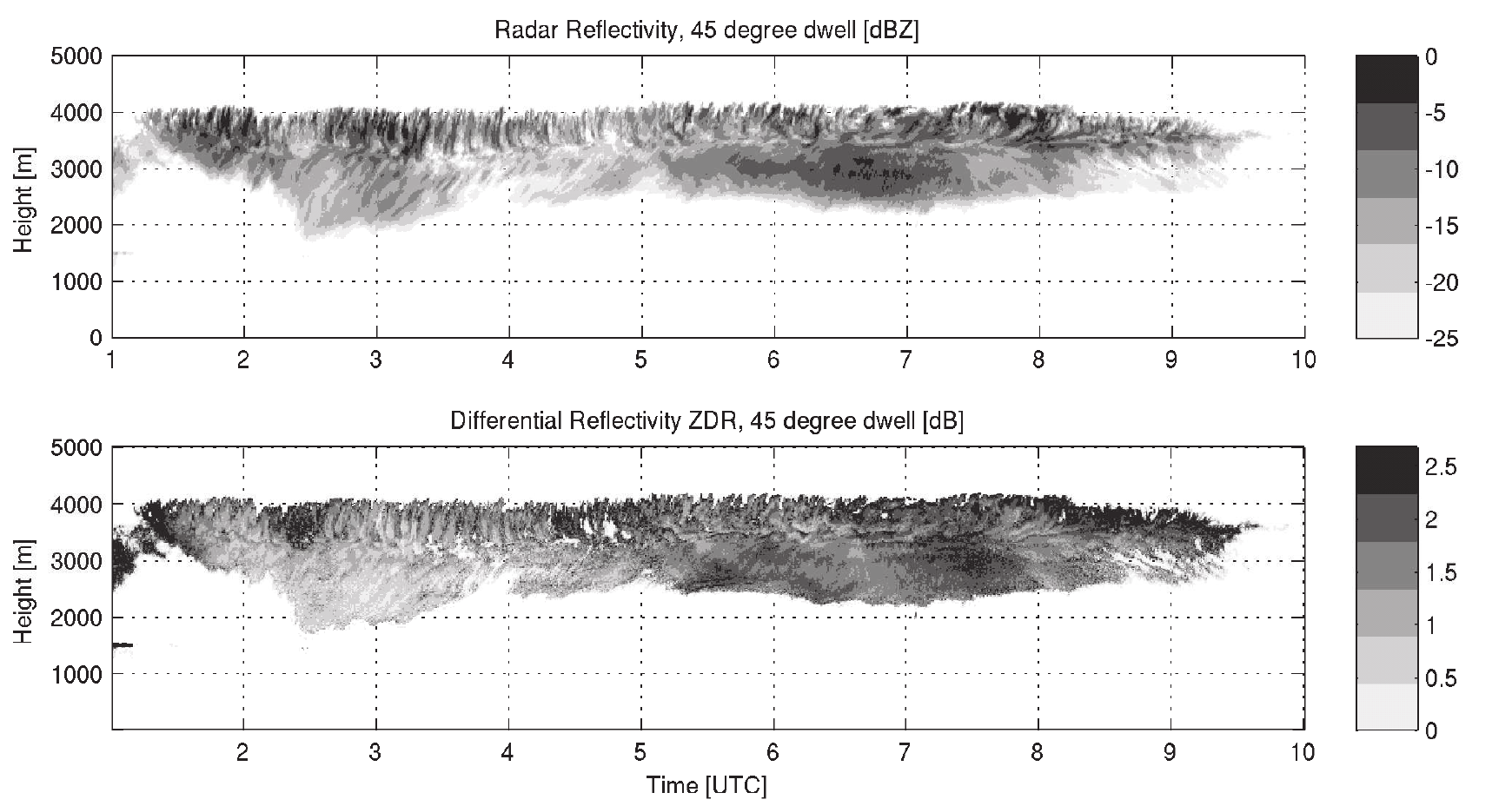}
\caption{\label{7lidars2}$45^{\circ}$ radar observations. Top panel shows the reflectivity recorded by the 35-GHz cloud radar; lower panel shows the differential reflectivity $Z_{DR}$. High values of $Z_{DR}$ correspond to returns from strongly oriented planar crystals.}
\end{figure*}

In addition to the vertical and near-vertical radar and lidar data shown in figure \ref{7lidars}, we also made simultaneous polarimetric radar measurements with the 35GHz cloud radar dwelling at $45^{\circ}$ (due West; the cloud is advecting in from the North East). The corresponding time series is shown in figure \ref{7lidars2}. The radar reflectivity structure is very similar to that recorded by the vertical 94GHz radar, giving us confidence that the cloud is essentially homogeneous within a 4km range of Chilbolton, and that we are sampling similar ice particles. Also shown in this figure is the differential reflectivity $Z_{DR}$. This parameter is the ratio of the co-polar radar reflectivity for horizontally polarised waves vs. those polarised at $45^{\circ}$ from horizontal (and perpendicular to the direction of propagation). Calculations (Appendix B) show that thin pristine crystals which are horizontally oriented produce a strong $Z_{DR}$ signature (up to 3.5~dB), whilst we expect randomly oriented crystals and low-density/irregular particles have $Z_{DR}\approx0$~dB. Crucially, if there are only a few pristine crystals mixed in with large irregular particles, the $Z_{DR}$ signature is masked (see Hogan \etal~2002,2003a; Field \etal~2004), because the reflectivity at both polarisations is dominated by the large irregular particles. 

\begin{figure}
\centering
\includegraphics[width=3in]{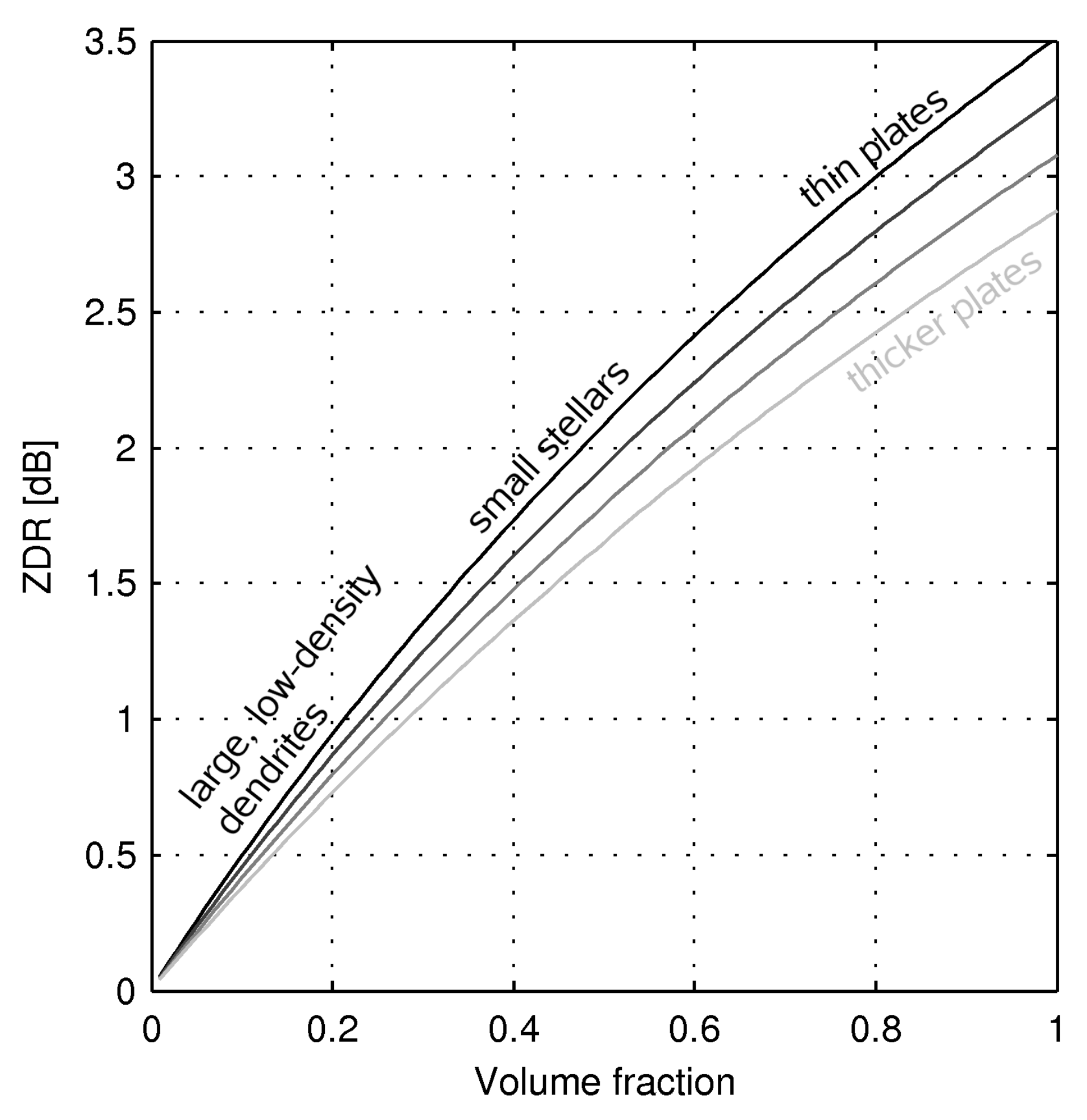}
\caption{\label{ZDR_fv}$Z_{DR}$ for horizontally oriented planar crystals observed at $45^{\circ}$ elevation as function of volume fraction. Lines indicate range of crystal aspect ratios observed by Ryan \etal~1976 (0.02--0.14 in steps of 0.04 from darkest to lightest). Typical ice/air volume fractions for some crystal types are indicated, inferred from Ryan \etal~and Heymsfield and Kajikawa (1987). }
\end{figure}

The observed values of $Z_{DR}$ demonstrate that the 15~dB of specular reflection seen in the lidar data is not the result of a few perfectly formed and oriented crystals, but reflects a broader population of planar crystals falling close to horizontal. Most of these are probably imperfect as a result of growing at water-saturation, and have a distribution of orientations centred around horizontal; but collectively these crystals have enough faceted surface normal to the lidar beam to produce a strong specular signal. Since crystals grown in this temperature range are typically very thin, $Z_{DR}$ is primarily sensitive to the crystal ice/air volume fraction (the volume of the ice crystal divided by a spheroid of the same overall dimensions), and calculated values for perfectly horizontal crystals are shown in figure \ref{ZDR_fv} . The detail of this calculation is described in Appendix B; we have marked on the parts of the curve corresponding to hexagonal plates and small stellar crystals as inferred from Ryan \etal's (1976) data. Given the cloud top temperature we anticipate growth initially of simple plates, evolving into branched stellar crystals after 1-2 minutes of growth, implying $Z_{DR}\approx$~2--3~dB; the observed differential reflectivity measurements seem to back up this expectation, with much of the virga having values in this range. There is a fascinating amount of structure in the $Z_{DR}$ observations however; presumably this is largely the result of size sorting in the fall streaks yielding streaks of high $Z_{DR}$ where the crystals are small and dense, and lower $Z_{DR}$ where the crystals are larger and more tenuous. The fall streak structure is similar to that observed in supercooled stratocumulus by Field \etal~(2004). We also note that the exact crystal habit is quite sensitive to the growth temperature, and given the range of temperatures spanned by the supercooled layer, it is quite likely a range of planar habits develop simultaneously, and are then sorted as they fall.
The very largest values of $Z_{DR}$ ($\approx3$dB) seem to be found only close to the very top of the cloud: presumably these are small plate-like crystals in the early stages of growth, with a volume fraction close to unity (see figure \ref{ZDR_fv}); the crystals lower down will likely have been growing for longer, developing tenuous branches, perhaps aggregating, and will therefore will have a lower volume fraction and lower $Z_{DR}$. The observation of strong differential reflectivity even at the very base of the virga where the crystals are evaporating is further evidence that the lack of specular reflection in the evaporation zone is due to the crystals becoming rounded (whilst still maintaining their horizontal orientation) rather than their orientation being disrupted, for example by turbulence associated with evaporating cooling.

\section{17 months of statistics}
We now present a statistical analysis of specular reflection from 17 months of continuous observations. A number of the studies cited in section 2 have calculated statistics on the occurrence of specular reflection, as summarised in table \ref{otherstudies}. Many of the results are inconsistent with one another, and all but Young \etal~(2000) and Hogan and Illingworth (2003) have effectively excluded profiles containing supercooled liquid. Unfortunately these latter two studies sampled quite a small number of clouds. It is therefore desirable to calculate some statistics of occurrence for specular reflection using our long continuous data set, and 
since we are able to range-resolve the position of the oriented crystals we can readily investigate their occurrence below supercooled layers, as well as in ice-only profiles. We note that our results are distinct from those derived from POLDER since our 30s$\times$36m sample volume is much smaller. The colour ratio and Doppler measurements allow us to explore more deeply some of the properties of these oriented ice crystals, such as how fast they fall, and what part of the distribution they likely represent.

 The data were analysed as follows. Boundary layer aerosol and precipitation were excluded by removing targets which produced a continuous profile of backscatter down to the lowest range gate. This effectively excludes supercooled stratocumulus from our analysis, and may also exclude occasional mid-level cloud where ice fall streaks extend into the boundary layer aerosol. In general however this procedure is quite effective at isolating non-precipitating cloud. As an additional condition we restrict our attention to cloud where the temperature is colder than $-2.5^{\circ}$C in order to exclude melting ice particles; the temperature is estimated from the Met Office's 12km forecast model output over Chilbolton (see Illingworth \etal~2007). 

We identify cloud as range gates where the backscatter is greater than $5\times10^{-7}\mathrm{m^{-1}sr^{-1}}$ from both lidars. The presence of supercooled liquid droplets is inferred by backscatter from the off-vertical ceilometer exceeding $10^{-4}\mathrm{m^{-1}sr^{-1}}$ and this range gate is flagged and removed. The backscatter from the three range gates below this are also removed from the analysis since these may also contain liquid, and all the data from range gates above is also discarded since  liquid cloud may well have significant structure on the scale of hundreds of metres and as a result there may be some differential attenuation between the zenith and off-zenith lidar beams, which could perhaps lead to an ambiguous interpretation of colour ratios measured from ice cloud above. The remaining ice cloud is divided into data with colour ratios $<0$~dB (specular reflection) and data with colour ratios $>0$~dB (`normal' ice cloud). Note that the lidar signal is attenuated even in glaciated clouds, and whilst it is often possible to obtain a complete backscatter profile through a relatively shallow cirrus cloud or altocumulus virga, for deep stratiform ice clouds we are typically only able to gather statistics from the lowest kilometre or two of cloud. Convective clouds are effectively excluded since the lidar is not able to see into the liquid cloud base. This analysis was performed for a total of 503 days (approximately 17 months) of continuous lidar observations between 12 September 2006 and 31 January 2008. 

\subsection{Occurrence by temperature}

\begin{figure}
\centering
\includegraphics[width=3in]{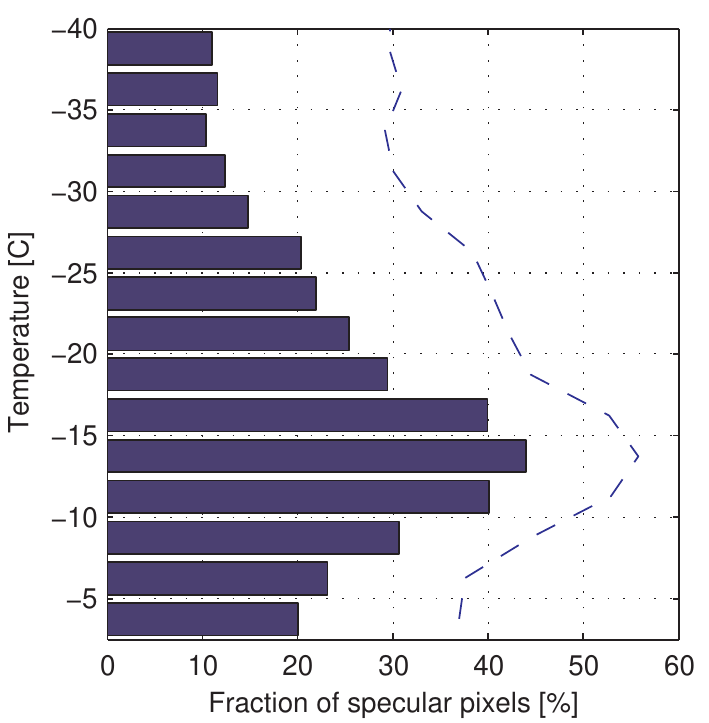}
\caption{\label{A1} Percentage of ice cloud which is identified as containing specularly reflecting ice crystals, binned as a function of temperature. A distinct peak is apparent close to $-15^{\circ}$C, where 40\% of ice cloud measured by both lidars is observed to be influenced by specular reflection. At temperatures colder than $-30^{\circ}$C this fraction is only 10\%. These figures are likely underestimates, and the dashed line shows the fraction of cloud affected if all colour ratios $<+3\mathrm{dB}$ are counted (see text). The total number of ice pixels in each temperature bin  ranged from $3\times10^4$ in the coldest bin, to $1.5\times10^5$ in the warmest. }
\end{figure}

Using the method described above we have calculated the fraction of cloud that is observed by both lidars and which is identified as containing specularly reflecting ice crystals. Overall we find that of the $1.55\times10^6$ ice cloud pixels analysed, 26\% were affected by specular reflection. Breaking this fraction down as a function of temperature reveals further insight, as shown by the horizontal bars in figure \ref{A1} - it is apparent that most of the specular reflection is concentrated at temperatures warmer than $-25^{\circ}$C, with a maximum in the region of $-15^{\circ}$C where over 40\% of ice cloud observed is affected by specular reflection. These observations seem consistent with the expectation that planar monocrystals typically grow in the temperature range $-9$ to $-22^{\circ}$C (Pruppacher and Klett 1997).
There is no obvious `gap' corresponding to the window of column/needle growth between $-3$ and $-8^{\circ}$C, but since there are few ice nuclei at these warm temperatures anyway, the specular in this region is likely planar crystals formed at colder temperatures sedimenting in from above. The maximum occurrence around $-15^{\circ}$C seems sensible, since thin planar crystals are well known to grow in this temperature region.

At temperatures between $-30$ and $-40^{\circ}$C only around $10$\% of cloud is affected by specular reflection. This is a significantly smaller fraction than observed in cirrus clouds by Thomas \etal~(1990), although their scanning lidar arrangement may be a more sensitive tool to detect weak specular reflection, since such a setup is not complicated by differential absorption effects (see below). 

Figure \ref{A4} shows the observed distribution of lidar colour ratios as a function of temperature. The lack of bimodality around the $0\mathrm{dB}$ mark suggests that in fact specular reflection is likely to be occurring weakly in some of the cloud where we measure positive colour ratios. This is a result of the increased absorption at $1.5\um$ compared to $905\nm$ as discussed in section 2. Specifically, this would correspond to cases where the specular reflection component of the backscatter is of comparable magnitude to the backscatter from the other ice particles in the cloud, and the overall colour ratio is positive (our off-zenith measurements indicated that ice particles which are not affected by specular reflection yield colour ratios in the range $+3$ to $+12\mathrm{dB})$. 

The distinction between specular and non-specular ice is most obvious at the mid-level temperatures between $-10$ and $-20^{\circ}$C, where a long straight tail extending out to $-20\mathrm{dB}$ is apparently distinct from a more bell shaped distribution for normal ice cloud, with the two curves intercepting at approximately $+3\mathrm{dB}$.  At temperatures colder than $-30^{\circ}$C the distribution is much narrower with relatively few colour ratios $<0\mathrm{dB}$, and the magnitude of the specular reflection (as characterised by the colour ratio) in these cold clouds is much weaker than at $-15^{\circ}$C. 


We note that if it were assumed that all values less than $+3\mathrm{dB}$ in fact corresponded to some specular reflection occurring in the lidar sample, the overall fraction of ice cloud affected would rise from 26\% to 40\%. The occurence by temperature following this assumption is shown by the dashed line in figure \ref{A1}. Our estimates of the frequency of specular reflection should therefore be considered as lower bounds on the true occurence, and this emphasises the large fraction of cloud in which specular reflection is possible, especially at the mid-levels. In the subsections that follow we will use the 0~dB threshold to identify cloud as specular or not, ie. where the effect is strong enough to be measureable without ambiguity.




\begin{figure}
\centering
\includegraphics[width=3in]{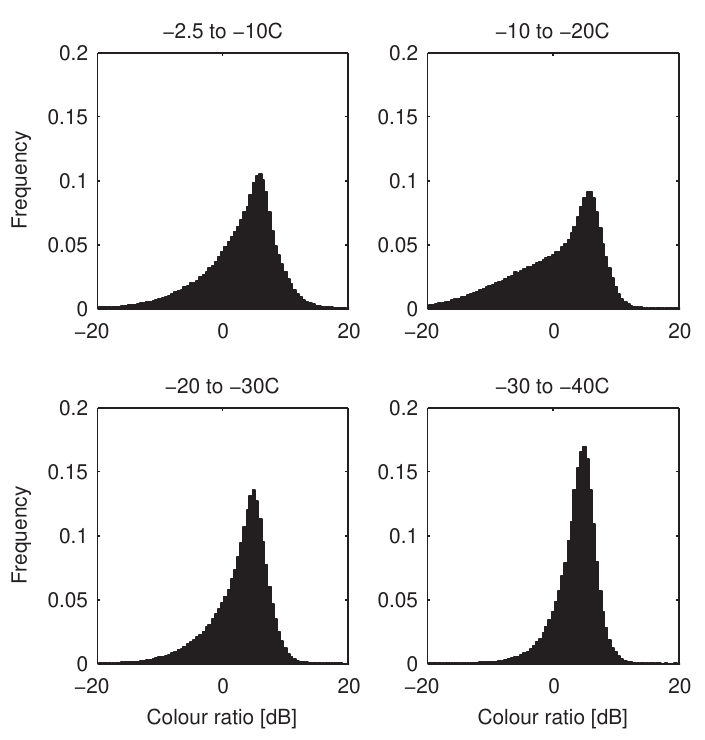}
\caption{\label{A4} PDFs of the lidar colour ratio seperated by temperature. At temperatures characteristic of planar growth (especially between $-10$ and $-20^{\circ}$C) the distribution has a long tail with ratios $<0\mathrm{dB}$ (specular reflection), whilst at colder temperatures the distribution is much narrower and concentrated at colour ratios $>0\mathrm{dB}$ (normal ice cloud).}
\end{figure}

\subsection{Fall speeds}

\begin{figure}
\centering
\includegraphics[width=2.75in]{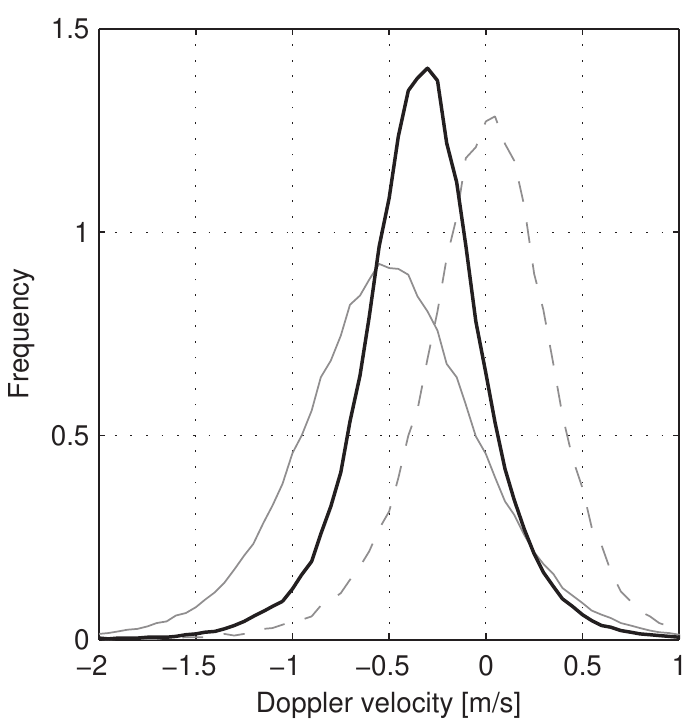}
\caption{\label{A2} Distribution of Doppler velocity for specularly reflecting ice crystals (black line),  `normal' ice cloud (solid grey line), and supercooled liquid droplets (dashed grey line). Frequencies normalised so area under the curve = 1. Mean values are $-0.32~\mathrm{ms^{-1}}$, $-0.49~\mathrm{ms^{-1}}$, and $0.00~\mathrm{ms^{-1}}$ respectively.}
\end{figure}

Applying the method of identification above, we can use the Doppler velocity measurements from the vertical lidar to study how fast the oriented crystals fall. The frequency distribution of Doppler velocities for all cloud affected by specular reflection is shown in figure \ref{A2}, which also shows the velocity distributions for normal ice cloud and liquid droplets. The frequencies are normalised so that the area under each curve is equal to 1. Note that the velocity distributions are broadened somewhat by the range of vertical air motions in the clouds, as well as by variability in the particle terminal velocities. The distribution for liquid droplets has a mean velocity of $0.00~\mathrm{ms^{-1}}$, the width of the distribution is likely dominated by turbulence of the kind observed in figure \ref{teste}. For normal ice cloud we measure a relatively broad distribution of velocities, with a mean value of  $-0.49\mathrm{ms^{-1}}$ and standard deviation of almost exactly the same magnitude.
This suggests that in the majority of ice clouds sampled by the lidar, the lidar signal is controlled by ice particles of several hundred microns in size. In cloud affected by specular reflection, the range of crystal fall speeds is rather narrower, with a mean value of $-0.32\mathrm{ms^{-1}}$. This is in keeping with the idea that specular reflection is produced by planar crystals from a relatively limited range of Reynolds number where the fall orientation is stable and the fall speeds are relatively slow. The corresponding Reynolds numbers for various assumed crystal habits have been calculated using Heymsfield and Kajikawa's (1987) relationships. Segregation of the data by colour ratio  (not shown for brevity) shows a slight decrease at the most extreme colour ratios (10-20\%); however this decrease is much smaller than the variation between the different assumed crystal habits ($\mathrm{Re}\approx5$ for hexagonal plates vs $\mathrm{Re}\approx20$ assuming stellars). In any case it seems that the specular backscatter and Doppler velocity is weighted to crystals falling with $\mathrm{Re}\sim10$, with a factor of 2 uncertainty. This is consistent with Sassen's light pillar observations, and the case study in section 4.

\begin{figure}
\centering
\includegraphics[width=3.25in]{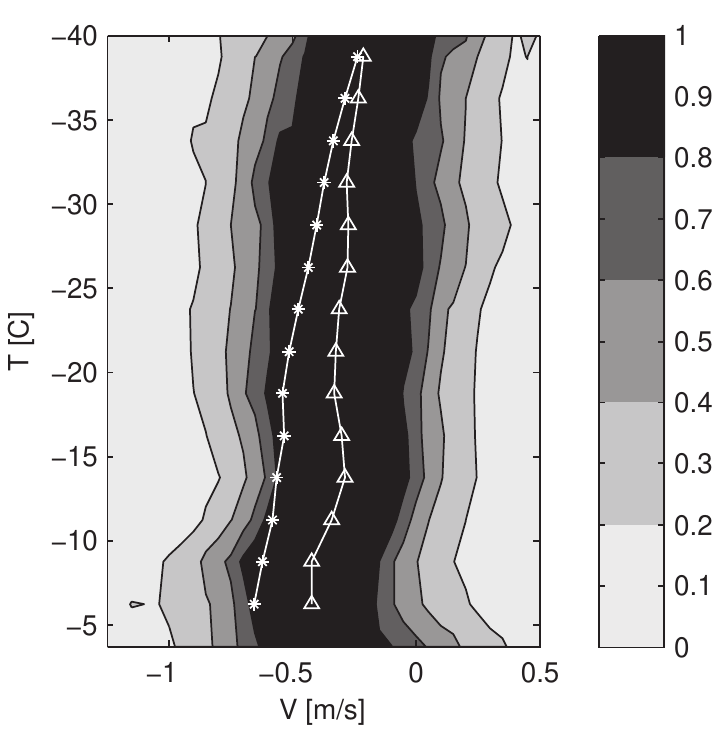}
\caption{\label{A6} Distribution of specular fall speeds as a function of temperature, filled contours indicate frequency normalised for each temperature bin in 0.2 [$\mathrm{sm^{-1}}$] intervals. White line with triangles indicates the mean values for specular cloud as a function of temperature; stars indicate same but for normal ice clouds.}
\end{figure}


Plotting the Doppler velocity distribution for specular crystals as a function of temperature yields figure \ref{A6}. The average particle fall speed for normal ice cloud was calculated by Westbrook and Illingworth (2009) and was found to increase systematically with temperature from $0.3\mathrm{ms^{-1}}$ at $-35^{\circ}$C, to $0.65\mathrm{ms^{-1}}$ at $-5^{\circ}$C, demonstrating the influence of larger particles and broader size spectra at warm temperatures. In contrast to this, the velocity distribution for specularly reflecting crystals shows a rather weaker variation with temperature, with average crystal fall speeds increasing from $0.25\mathrm{ms^{-1}}$ at $-35^{\circ}$C, to $0.3\mathrm{ms^{-1}}$ at $-15^{\circ}$C; there appears to be a slight broadening of the distribution for the warmest clouds, with an average velocity of $0.4\mathrm{ms}^{-1}$ at $-5^{\circ}$C. Possibly this corresponds to planar crystals formed at the mid-levels which thicken as they fall into air temperatures characteristic of columnar growth, increasing their weight relative to their drag, and allowing them to sediment more quickly for a given Reynolds number. As in section 4 the mean Doppler velocities and implied crystal Reynolds numbers are lower than the maximum stable values found by Kajikawa (1992) for all assumed crystal types.

\subsection{Crystal alignment}

\begin{figure}
\centering
\includegraphics[width=3in]{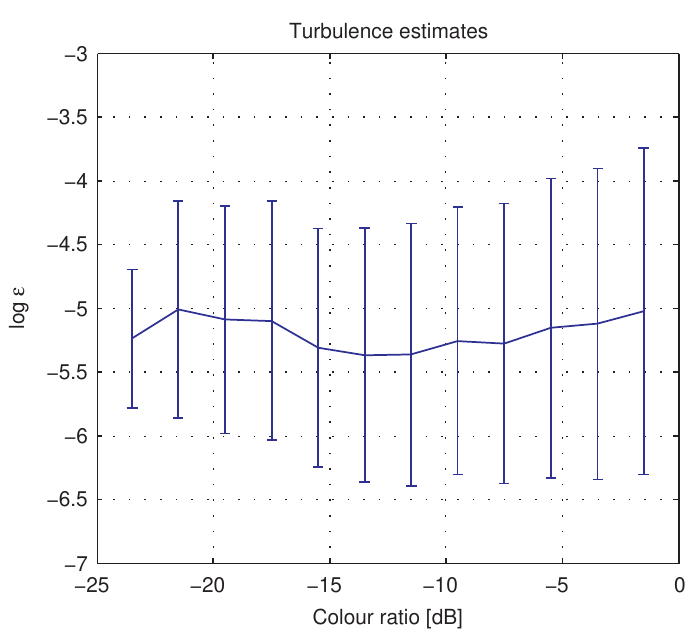}
\caption{\label{C9} Effect of turbulence on crystal orientation: line shows mean of $\log_{10}\epsilon$ measured by the radar as a function of colour ratio; error bar shows 1 standard deviation in $\log_{10}\epsilon$. }
\end{figure}

The correlation between turbulent air motion and colour ratio measurements has also been investigated, to determine whether the horizontal orientation of the ice crystals is affected by the turbulent eddies. The eddy dissipation rate $\varepsilon$ was estimated as described in section 4, and the results were binned by colour ratio and dissipation rate. The resulting mean and standard deviation of $\log_{10}\epsilon$ is shown in figure \ref{C9}, and we observe that there is essentially no correlation between colour ratio and  turbulence. There is a hint that the standard deviation is slightly lower for the most extreme colour ratios; however this is because there are very few samples here. These statistics are consistent with the case study in section 4. Klett (1995) and Br\'{e}on and Dubrulle (2004) performed theoretical calculations of the influence on turbulence on crystal orientation, and these also indicate that for the observed levels of $\varepsilon$, crystals can be oriented to well within $1^{\circ}$ of horizontal. Our observed range of $\varepsilon$ values are consistent with other ice cloud observations (eg. Gultepe and Starr 1995, Bouniol \etal~2003). We note that the bulk of our $\epsilon$ values are an order of magnitude lower than in Klett's calculation for `weak' turbulence, and suggest that in most stratiform clouds, the ability of turbulent eddies to influence the air flow on the scales relevant to ice crystals is probably minimal. 

The lack of correlation observed above leads us to believe that the colour ratio is primarily sensitive to the fraction of ice particles which are pristine planar crystals with $\mathrm{Re}$ close to 10. 

\subsection{Supercooled layers}
\begin{figure}
\centering
\includegraphics[width=3in]{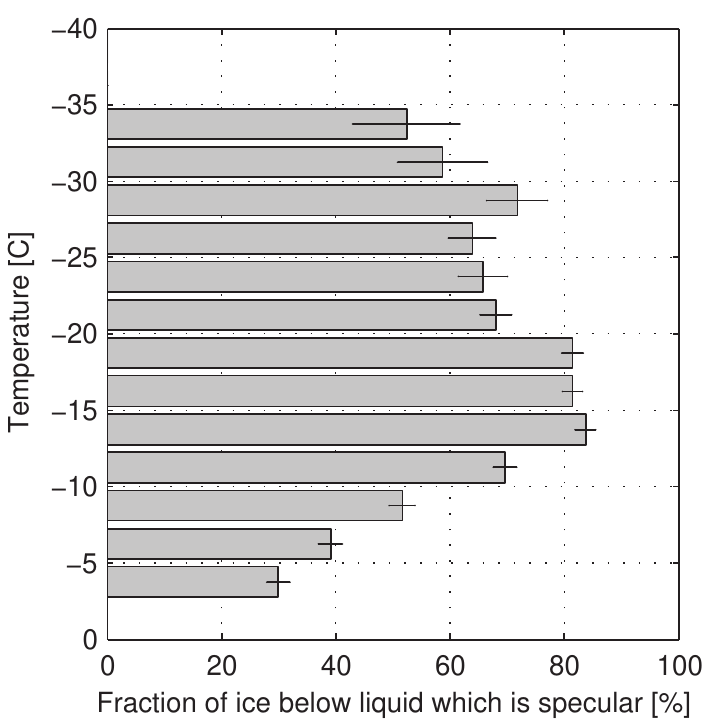}
\caption{\label{C10} Fraction of ice cloud observed below supercooled liquid which contains some specular reflection, binned as a function of the temperature of the liquid cloud. Horizontal lines indicate 95\% confidence intervals for these fractions (assuming that if multiple samples were made from the same underlying population, the fractions derived from those samples are normally distributed).}
\end{figure}

As mentioned in section 3, it is our experience that supercooled liquid layers are a frequent source of ice crystals producing specular reflection. Young \etal~(2000), Platt (1977), Sassen (1984) also observed specular ice crystals falling beneath supercooled liquid layers. We have investigated this link in a more quantitative manner by isolating liquid cloud using the off-vertical ceilometer in the manner described above, and then picking out the occasions when ice crysals are observed falling directly below this liquid layer. By the nature of the algorithm, we only pick out the lowest supercooled layer in multi-layered cloud systems. The fraction of these profiles affected by specular reflection is plotted in figure \ref{C10} as a function of the temperature of the liquid layer. It is immediately apparent that specular reflection is very frequent in these cloud types, with 80\% of ice falling below supercooled layers between $-12$ and $-20^{\circ}$C affected. This is consistent with in-situ observations by Locatelli \etal~(1983), Hogan \etal~(2003a) and Field \etal~(2004) who all observed pristine vapour-grown planar crystals below supercooled layers in this temperature range; it is also consistent with our case study in section 4. Ryan \etal~(1976) found all crystals grown at these temperatures to be simple planar types. Such crystals therefore indeed seem to be characteristic of mid-level supercooled clouds as suggested by Hogan \etal~and Field \etal~-- this may be a useful signature for the possible presence of supercooled liquid water in optically thicker clouds where the liquid cannot be seen by the lidar, but the specular reflection from the ice crystals falling out is readily observed. 

There is a sharp decrease in the fraction of specular profiles at temperatures warmer than around $-10^{\circ}$C; any crystals nucleated in such conditions are likely to be columns and needles, although there is still specular reflection observed in a significant fraction of cases. The lack of ice nuclei means that many of these warmer supercooled layers which are observed to be precipitating ice crystals are likely to be seeded from colder clouds above, in a similar way to the liquid layers embedded in the ice virga in fig \ref{7lidars}. At temperatures colder than $\approx-35^{\circ}$C there are no cases in our 17 month data set, which seems sensible as this is close to the limit for homogeneous freezing.

\begin{figure}
\centering
\includegraphics[width=3in]{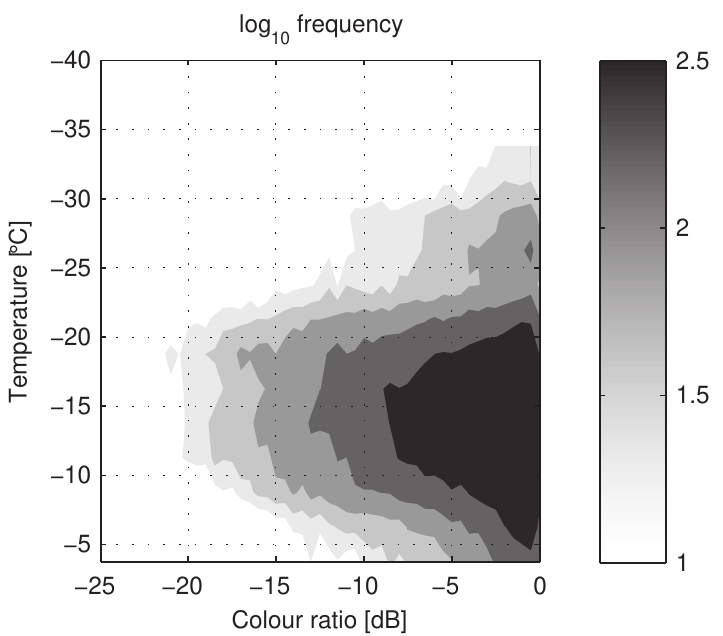}
\caption{\label{testcolour} Filled contour plot showing distribution of colour ratios in specularly reflecting ice cloud falling beneath supercooled liquid layers. Grey scale is $\log_{10}\mathrm{(frequency)}$ in the $2.5^{\circ}\mathrm{C}\times0.5\mathrm{dB}$ bins. Contours range from 1.0 to 2.5 in steps of 0.3.}
\end{figure}

Even supercooled layers at cold temperatures $<-22^{\circ}$C often have specular reflection in the crystals falling below, around two-thirds of the time. This conclusion is also supported by Heymsfield \etal's (1991) observation of a highly supercooled altocumulus layer ($\approx-30^{\circ}$C) with specular ice crystals precipitating out. To investigate this behaviour in more detail, figure \ref{testcolour} shows the distribution of colour ratios measured in the specular ice, as a function of supercooled layer temperature. This distribution demonstrates that specular reflection is strongest in the ice falling from supercooled layers which lie between $-10$ and $-20^{\circ}$C (where the droplets will freeze into planar monocrystals eg. Ryan \etal~1976). At temperatures below $-20^{\circ}$C, the magnitude of the colour ratio decreases substantially, indicating that planar monocrystals are typically a minority in these colder clouds. In what follows, we speculate on the origin of the specular crystals which \emph{are} present in cold clouds and why they are fewer in number than at warmer temperatures, by linking previous laboratory and observational measurements of crystal growth to the data in figures \ref{C10} and \ref{testcolour}.

Growth at temperatures $<-22^{\circ}$C has traditionally been associated with columnar crystal habits (Pruppacher and Klett 1997), and one may hypothesise that the observed specular reflection is the result of Parry-oriented columns (major axis horizontal, prism face horizontal), rather than plates; such a case has been documented by Sassen and Takano (2000). However observations of the associated Parry arc are very rare, and Sassen and Benson (2001) have argued that in general only planar ice crystals will produce specular reflection.

Perhaps a more likely explaination of our observations is that planar crystal growth is occurring at colder temperatures than expected. Planar crystals have been observed to grow at cold temperatures in a number of studies.
Aufm Kampe \etal~(1951) grew mainly plate-like polycrystals from a supercooled water cloud in a cold chamber at $-22^{\circ}$C and $-29^{\circ}$C, along with a smaller fraction of simple plates and columns. Bailey and Hallett (2002,2004) nucleated and grew ice crystals on kaolinite and bare glass fibres, and found that plate-like growth in mono- and poly-crystalline forms dominated down to approximately $-38^{\circ}$C. Bacon \etal~(2003) nucleated and observed the very early stages of growth of supercooled droplets suspended in an electrodynamic trap and observed plate-like growth at temperatures as cold as $-38^{\circ}$C. Fleishauer \etal~(2002) observed a mixed-phase cloud topped at $-30^{\circ}$C: the ice particle images from that case study seem to show a mixture of planar mono- and poly-crystals. Sakurai (1968) observed ice crystals in supercooled fogs between $-20$ and $-28^{\circ}$C and found irregular polycrystals, thick plates and short (almost 1:1 aspect ratio) columns. These laboratory and observational data suggest that crystals nucleated and grown from droplets at these cold temperatures often favour plate-like growth, albeit typically as a polycrystal rather than simple monocrystalline plates.

Pitter and Pruppacher (1973) have investigated the freezing of fairly large liquid water droplets ($180$--$800\mu$m in diameter) by immersion and contact nucleation (using clay particles), and observed that for a given supercooling $\Delta T$, there was a critical size for the droplet to grow into a polycrystal: $D_{c}\propto(\Delta T)^{-3}$. Bacon \etal~(2003) found that their $40\um$ diameter droplets grew into polycrystals at temperatures colder than $\approx-22^{\circ}$C. Based on this, one might expect a smaller $20\um$ droplet to develop into a polycrystal at $\approx-27^{\circ}$C; for a $10\um$ droplet, monocrystals would dominate down to $\approx-35^{\circ}$C. These thresholds are approximately consistent with the extrapolation of Pitter and Pruppacher's data to cloud droplet sizes. 

We suggest therefore that our observed specular reflection in cold clouds is the result of small cloud droplets freezing into simple (thick) plates which produce specular reflection, whilst the larger droplets in the droplet distribution freeze into irregular polycrystals. Whereas in clouds warmer than $-20^{\circ}$C the vast majority of cloud droplets will freeze as monocrystals irrespective of droplet size, for colder clouds an increasing fraction of the droplets frozen will be polycrystalline, and incapable of producing specular reflection. 

We note that although on average specular reflection in cold clouds is much weaker, there are still \textit{individual cases} where it is large ($\approx-10$dB) - these may be examples where the droplets are particularly small. For the low liquid water content stratiform clouds which dominate our observations, we anticipate there may be a stronger weighting towards smaller droplets (and therefore more single crystals) than in more convective-type clouds. Note that the droplets in Heymsfield \etal~'s (1991) altocumulus layer with specular reflection in the virga below had a mean diameter of only $\approx10\mu$m.

\subsection{Layers of planar crystals}
\begin{figure}
\centering
\includegraphics[width=2.75in]{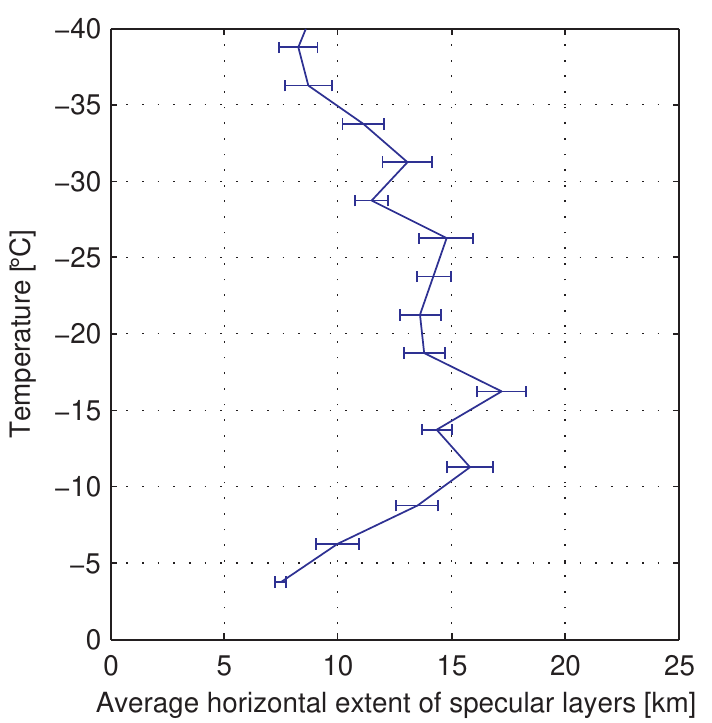}
\caption{\label{cellsize} Mean horizontal extent of specularly reflecting layers, as a function of the average layer temperature. Error bars indicate uncertainty in the mean values.}
\end{figure}

Our observations (eg. section 3) indicate that specular crystals form widespread layers stretching over many kilometres; this also fits in with previous lidar observations (Platt 1977). Aircraft observations by Korolev \etal~(2000) suggest that branched planar crystals often occur in `cells'  many kilometres across. To get an idea of the typical scale of these planar crystal layers, we have identified areas of cloud which have at least one common range gate with a negative colour ratio in two consecutive vertical profiles - in this way we are able to identify continuous layers of specular reflection, and characterise their horizontal extent. Of course there is the complication that the lidar is frequently blocked by low level liquid clouds. To ameliorate this effect (which will tend to bias our estimates too low) we allow gaps of up to 5 minutes in the layer so long as there is still an adjacent region of specular reflection on both sides of the gap. Using the wind speeds from the Met Office forecast model, figure \ref{cellsize} shows the mean horizontal extent of the planar crystal layers as a function of the average temperature of the layer. It is apparent that in mid-level clouds the average extent is approximately 15~km, whilst at colder temperatures the layers appears to be less long-lived with an average extent of approximately half that value. In light of the discussion above regarding specular reflection below liquid layers, this may perhaps be the result of the more transient nature of supercooled liquid clouds at these cold temperatures, which typically are only $\sim5$~km in width at $-35^{\circ}$C (Hogan \etal~2003b). 
There is a also a systematic decrease in layer extent at warm temperatures in figure \ref{cellsize}. The reason for this is unclear, although it may represent a shift in weighting of the statistics from widespread supercooled altocumulus virga to the base of deeper ice clouds associated with approaching fronts. We note that at all temperatures the distribution of layer widths is quite broad, with some cases lasting for hours on end, whilst others only occur in isolated pockets or fall streaks, and we observed layers from $\sim1$~km up to more than $100$~km in extent. This large variability is also consistent with Korolev \etal's observations. 

The depth of the specular layers has also been characterised - we estimate a mean depth of 600m, approximately independent of temperature. This figure is a convolution of the true layer depth and the extinction of the lidar beam in optically thick clouds, and is likely an underestimate.

\subsection{Backscatter and radar reflectivity}

\begin{figure}
\centering
\includegraphics[width=3in]{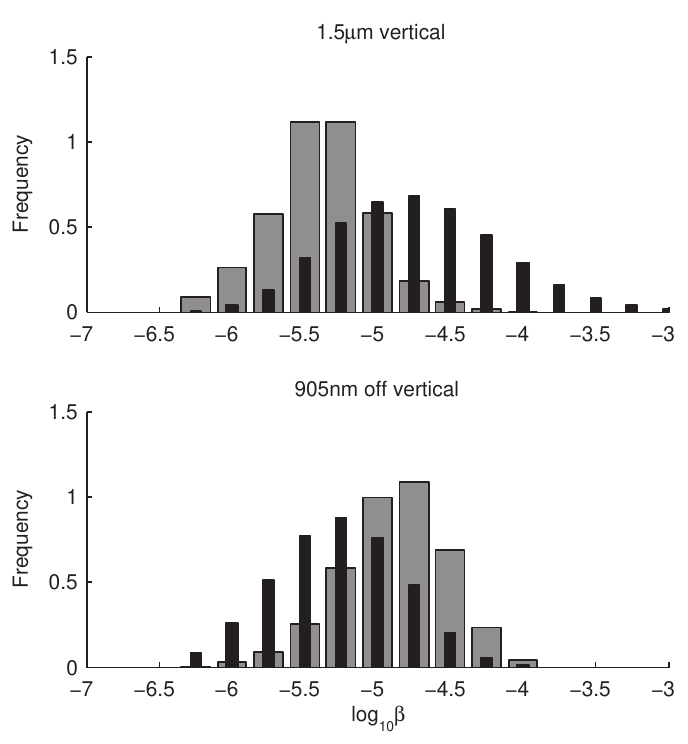}
\caption{\label{A4A} PDFs of lidar backscatter recorded by the vertical $1.5~\mu$m and off-vertical 905~nm lidars. Black bars are the PDFs for cloud identified as containing specularly reflecting ice crystals, grey bars are PDFs for normal ice cloud.}
\end{figure}

Figure \ref{A4A} shows distributions of attenuated backscatter recorded from each lidar, with separate bars indicating specular and normal ice cloud. The data from the vertically pointing $1.5\um$ lidar shows that the backscatter from clouds affected by specular reflection is typically a factor of $5\mathrm{dB}$ larger than that recorded for normal clouds. From the $905\nm$ data we observe the opposite, ie. clouds where specular reflection is present typically have a backscatter which is typically around $5\mathrm{dB}$ lower than that recorded in normal ice cloud. This suggests that cloud where specular reflection is occurring may tend to be optically thinner than average. However, the interpretation of this result is subtle since the lidar is attenuated as it passes through the cloud, and our experience suggests that specular reflection is rarely observed at cloud base (because the crystal facets become rounded as they evaporate), which may bias the statistics towards lower values of off-zenith backscatter. 

\begin{figure}
\centering
\includegraphics[width=3in]{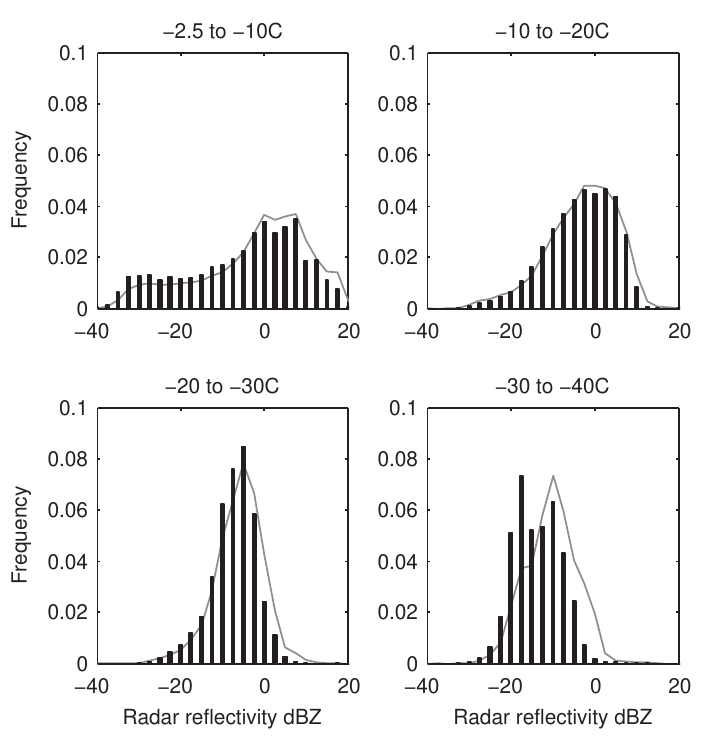}
\caption{\label{C8} PDFs of radar reflectivity for clouds containing specularly reflecting crystals (black bars), all ice cloud detected by both lidars (grey lines).}
\end{figure}

Radar reflectivities have been measured from the 35GHz cloud radar at Chilbolton, and data were analysed for the 7 month period June 2007 to January 2008. Figure \ref{C8} shows the distribution of radar reflectivity measured in ice clouds which are also observed by both lidars, and it appears that the reflectivity distribution in clouds which are affected by specular reflection (black bars) is almost identical to the reflectivity distribution for all ice cloud (grey lines). The suggestion then is that the ice water content in cloud where there is specular reflection is representative of the complete subset of ice cloud observed by both lidars. 

\section{Summary}
`Mirror-like' specular reflections from horizontally oriented planar ice crystals have been observed by comparing measurements from a vertically pointing Doppler lidar and a second lidar pointing slightly off-zenith. This comparison allows supercooled liquid cloud (which is highly reflective at both elevations) to be separated from layers of oriented planar crystals (which only produce a strong signature at vertical). Whilst most previous studies of specular reflection have focussed on glaciated cloud profiles, we find that strong specular reflection is most common in mid-level mixed-phase clouds, and is observed 85\% of the time in ice falling from supercooled layers at $-15^{\circ}$C. These crystals typically formed well defined layers with a mean extent of 15km, although numerous cases 100km or more across have been observed. Similar vast regions of planar ice crystals have been observed beneath supercooled boundary-layer clouds in Alaska (A. Korolev, personal communication).

A case study of a long-lived supercooled layer with planar ice crystals precipitating below was presented as a typical scenario where strong specular reflection is likely to be observed, and the backscatter at vertical was a factor of 30 (15dB) larger than observed off-zenith. Simultaneous polarimetric radar measurements confirmed that the general crystal population was well oriented with values of $Z_{DR}$ typically $\approx2$dB. Observations at the base of the ice virga reveal that specular reflection does not occur in this region, even though the $Z_{DR}$ measurements show that the crystals are still well oriented, and this is evidence that the crystals are evaporating rapidly and becoming rounded in shape. This may be a useful observational technique to identify whether crystals are evaporating layer-by-layer at small subsaturations, or whether the evaporation rate is rapid and uncontrolled in drier air (Nelson 1998, Hallett \etal~2001).

High temporal resolution Doppler radar measurements reveal that there was appreciable convective overturning at cloud top due to radiative cooling of the liquid layer to space, and this turbulence extended down into the ice virga. Despite this turbulent air motion ($\epsilon\approx3\times10^{-4}\mathrm{m^{2}s^{-3}}$), strong specular reflection was still observed in this region of virga, indicating that turbulence does not have a strong effect on crystal orientation in stratiform clouds. This is backed up by the lack of correlation between the measured colour ratio and $\epsilon$ in section 5.3.

Lidar Doppler measurements indicate that these horizontally oriented crystals fall slowly, with a mean velocity of $\approx0.3\mathrm{ms^{-1}}$ and a weak temperature dependence, in contrast to normal ice cloud where the observed sedimentation velocities in warm clouds are twice as fast as those at $-40^{\circ}$C. Using velocity-diameter relationships, we estimate that the most strongly oriented planar crystals have a Reynolds number in the region of $\mathrm{Re}\sim10$ with a factor of 2 uncertainty. This value is consistent with previous measurements of light pillars by Sassen (1980). The associated crystal dimensions are approximately 400--1200$\mu$m depending on the assumed crystal habit. 

Future work will focus on the analysis of data from a depolarisation channel added to the Doppler lidar in 2008, providing  high resolution measurements of depolarised backscatter and depolarised Doppler velocity. 

Interestingly, significant specular reflection is observed in ice falling below supercooled liquid layers at temperatures between $-22$ and $-35^{\circ}$C. Freezing of large cloud droplets at such temperatures typically produces irregular plate-like polycrystals; however polycrystal development is a strong function of droplet size, and we suggest that our observations can be explained by freezing of relatively small droplets, which develop into simple thick plates and produce specular reflection. This may be more favourable in the gently forced stratiform clouds that we are sampling; in more convective clouds with bigger droplets polycrystals would almost certainly dominate.

Finally, whilst we have observed that specular reflection is strong in many cases, we see from figure \ref{A4} that in the majority of cloud it occurs at a relatively weak level with colour ratios close to 0~dB, and indeed at all temperatures the modal colour ratio is positive (ie. no specular reflection). This lends support to the in-situ observations of Korolev \etal~(2000) which indicate that most ice particles in thick stratiform clouds are `irregular', probably a mixture of complex polycrystals and aggregated particles (Stoelinga \etal~2007), but with few planar monocrystals mixed in. The important exception to this rule appears to be thin supercooled layer clouds, particularly those in the range $-10$ to $-20^{\circ}$C, which our measurements indicate are a prolific source of pristine planar monocrystals.

\ack We are grateful to the staff at STFC Chilbolton for their hard work in operating and maintaining the lidars and radars. CDW acknowledges helpful correspondence with Alexei Korolev (Environment Canada) and John Hallett (DRI). This work was funded by the Natural Environment Research Council grants NER/Z/S/2003/00643 and NE/EO11241/1. 

\section*{Appendix A: Calibration of the 1.5$\mu$m Doppler lidar}
\begin{figure}
\centering
\includegraphics[width=3in]{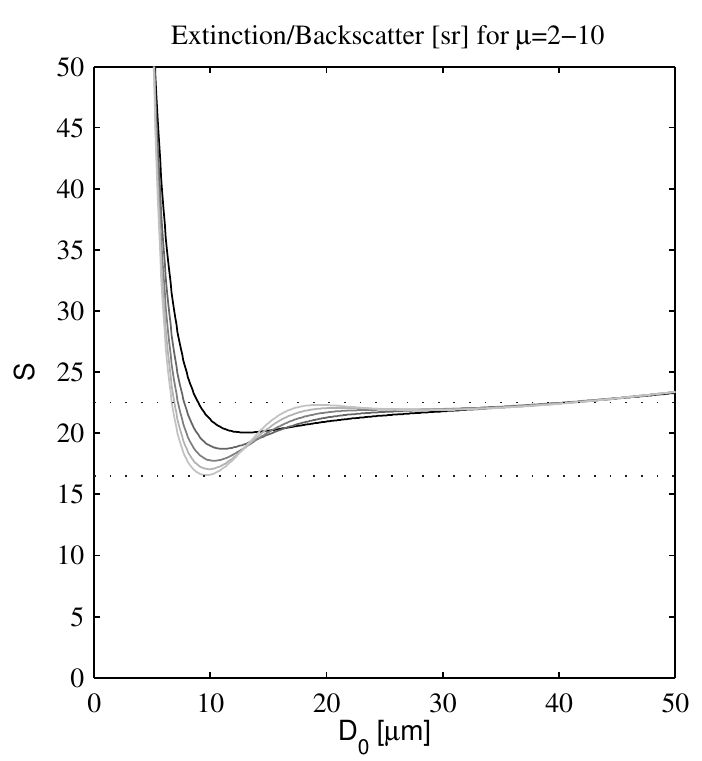}
\caption{\label{calibration} Extinction to backscatter ratio for a gamma distribution of liquid droplets with $\mu=2$ (darkest gray) to $10$ (lightest gray).}
\end{figure}

We follow O'Connor \etal~(2004) and compute the extinction to backscatter ratio $S$ for a gamma distribution of liquid water droplets for the new lidar wavelength of $1.5\um$. This is plotted as a function of the median volume droplet diameter $D_0$ for various values of the dispersion parameter $\mu$ in figure \ref{calibration}, where the droplet diameter distribution takes the form $n(D)\propto(D/D_0)^{\mu}\exp[-(3.68+\mu)D/D_0]$. For realistic values of the median volume diameter $D_0$ (8--40$\um$) and $\mu$ (2--10), $S=20\pm20\%$. For an optically thick liquid cloud which completely attenuates the lidar signal the integrated backscatter through the cloud is simply $(2S)^{-1}$ (see O'Connor \etal) and the calibration is applied simply by scaling the backscatter so that the integrated backscatter measured in thick stratocumulus cloud matches $S=20$. Because of the very narrow field of view, multiple scattering is neglected.

\section*{Appendix B: Differential reflectivity calculations}
We estimate $Z_{DR}$ at $45^{\circ}$ elevation for horizontal planar crystals by approximating them as oblate spheroids composed of a homogeneous mixture of air and ice inclusions. The incident wave is alternately transmitted in two orthogonal polarisations: one parallel to the horizontal plane (`H'), and a second at $45^{\circ}$ from the horizontal plane (`V'). Both are orthogonal to the direction of propagation and have amplitude $E_0$. Resolving the incident wave along the spheroid's major and minor axes the amplitudes are:
\begin{eqnarray}
E_0 &-& \textrm{ along major axis 1 (parallel to H)}\\
E_0/\surd2 &-& \textrm{along major axis 2 ($\perp$ to H)}\\
E_0/\surd2 &-&\textrm{along minor axis}
\end{eqnarray}
inducing dipole moments:
\begin{eqnarray}
E_0\alpha_{\mathrm{major}} &-& \textrm{ in the H channel}\\
0.5E_0(\alpha_{\mathrm{major}}+\alpha_{\mathrm{minor}}) &-& \textrm{ in the V channel}
\end{eqnarray}
where $\alpha_{i=\mathrm{major,minor}}$ are the polarisabilities:
\begin{equation}
\alpha_i=\textrm{particle volume}\times\frac{\epsilon-1}{1+L_i(\epsilon-1)}.
\end{equation}
The geometric factors $L_i$ for spheroids are detailed in Bohren and Huffman (1983); the permittivity of the air ice mixture is calculated using Maxwell-Garnett's rule for spherical inclusions (see Bohren and Huffman) - this prescription depends only on the volume fraction of ice in the mixture $f_v=(\textrm{volume of ice})/(\textrm{volume of spheroid})$. The differential reflectivity is then simply:
\begin{equation}
Z_{DR}=10\log_{10}\left[4\left(\frac{\alpha_{\mathrm{major}}}{\alpha_{\mathrm{major}}+\alpha_{\mathrm{minor}}}\right)^2\right]
\end{equation}
which is maximised for thin plates of solid ice ($f_v=1$) at $Z_{DR}=3.6$dB. Hogan \etal~(2002) have performed more extensive calculations of this parameter for $0^{\circ}$ elevation.

\end{document}